\documentclass[twocolumn,amsmath,amssymb]{revtex4}
\usepackage{graphicx}
\usepackage{dcolumn}
\usepackage{bm}
\usepackage{amsmath}

\newcommand{\be}{\begin{equation}}
\newcommand{\ee}{\end{equation}}
\newcommand{\ba}{\begin{eqnarray}}
\newcommand{\ea}{\end{eqnarray}}
\newcommand{\bal}{\begin{align}}
\newcommand{\eal}{\end{align}}
\newcommand{\n}[1]{\label{#1}}

\newcommand{\eq}[1]{(\ref{#1})}
\begin{document}
\title{Magnetic Field Evolution in Binary Neutron Stars}
\author{Shohreh Abdolrahimi}
\email{sabdolra@phys.ualberta.ca}
\affiliation{Theoretical Physics Institute, University of Alberta, 
Edmonton, AB, Canada,  T6G 2G7}
\date{\today}
\begin{abstract}
In this paper we investigate the  evolution of binary neutron stars, namely, their  magnetic field, spin, and orbital evolution. The core of a neutron star is considered to be a superfluid, superconductor type II. Flux expulsion of the magnetic field out of the core of a single neutron star has been discussed by previous authors. However, the evolution of the core magnetic field is substantially different for a binary neutron star. While for a single neutron star the fluxoids of the proton superconductor always move outward through the core, in a binary neutron star in accretion phase fluxoids move back into the core. The subsequent increase of the core magnetic field results in the increase of the surface magnetic field. We ask whether within the framework of this model the formation of millisecond pulsars (MSPs) is possible. We show that despite the increase of the core magnetic field, MSPs are formed in this model. The evolution of neutron stars with various orbital periods, magnetic fields, spin periods, and other parameters are numerically investigated.  The equation of state of the neutron star, initial orbital period of the binary, and other parameters of the binary have substantial effects on the evolution of period vs. magnetic field.
\end{abstract}
\maketitle
\section{\label{sec:level1}INTRODUCTION}
The evolution of the magnetic field of neutron stars has been widely discussed in many papers  \cite{Sir}-\cite{Ur}. The present study is mainly addressed to the coupled spin, magnetic field, and orbital evolution of a binary neutron star. In this paper we investigate  motion of fluxoids into and out of the  superfluid, superconductive core of a neutron star and the subsequent evolution of the magnetic field in the crust.  To derive the magnetic out-flux and in-flux rates, we consider various forces which act on fluxoids in the interior of a neutron star \cite{Din} and \cite{Gep2}.
The evolution of a binary star is affected by accretion from the stellar wind matter of its companion by various effects. 
These effects were discussed in \cite{Ur} under the assumption that magnetic field is confined within the  solid crust of a neutron star. In this paper we extend the results of \cite{Gep2} (for isolated neutron stars) to binary neutron stars, taking into account the various effects of the accretion discussed in \cite{Ur} and \cite{Ja1}. We show that accretion significantly modifies the evolution of the magnetic field. Namely, within the framework of this model, the magnetic field of a binary neutron star can increase, while magnetic field of a single neutron star always decreases. 

As a neutron star evolves, the strength of its surface magnetic field and spin period change. 
The $P_s-B_s$ diagram is usually used to present  distribution of pulsars of different surface magnetic fields $B_s$ and spin periods $P_s$ (here `s` stands for spin, not surface), or their evolution. Observed pulsars are distributed within two separate regions in $P_s-B_s$ diagram. MSPs are fast spinning stars ($P_s<10-20$ ms). Their surface magnetic fields $B_s$ is less than $10^{10}$ G. Other pulsars (normal pulsars) have higher magnetic fields and longer spin periods. Almost $50\%$ of MSPs are in binaries, while only $2\%$ of normal pulsars are found in binaries.  It is believed that newly born neutron stars have high surface magnetic fields and short spin periods. The spin period of single neutron stars 
increases during time, and their magnetic field decays. It has been suggested that MSPs are old binary neutron stars. According to the standard picture \cite{Pri}, \cite{Ill}, binary neutron stars pass through four evolutionary phases: the ejector, the propeller, the accretion from the wind of the companion, 
and the accretion resulting from the Roche-Lobe overflow. This four-phase model explains the low magnetic field and fast spin period of the millisecond pulsars,  the so-called recycled binary neutron stars. 

The mechanism of generation of the magnetic field of neutron stars is not totally understood. However, it has been suggested that magnetic field might penetrate through both the core and the crust of a neutron star \cite{Th}. However, as we have mentioned in our model, the evolution of the core magnetic field and subsequently the surface magnetic field are significantly different for binary and single neutron stars.
 In a binary neutron star, the magnetic field increases during the accretion phase due to the in-flux of the magnetic field into the core. This raises the intriguing question whether MSPs can still be formed within the framework of such models. We shall illustrate that if one includes the core magnetic field, there still exists the possibility for the formation of MSPs in binaries. However, not all initial values of the core magnetic field lead to the formation of MSPs.

As it was discussed in \cite{Ja1}, the orbital separation of a binary affects the rate of change its neutron star spin period. However, the orbital separation also evolves during the evolution.  We present the importance of the consideration of the orbital evolution on the final stage of a neutron star.

The decay rate of the magnetic field in the crust depends on the crustal conductivity. The conductivity depends on the equation of state (in the crust) and on the temperature, and hence, on the age of the neutron star. 
Therefore, the equation of state in the crust affects the evolution of a neutron star in the $P_s-B_s$ diagram. 
According to \cite{Ur}, the evolution of neutron stars with a stiff equation of state like the Friedman-Pardharipande are in agreement with the $P_s-B_s$ distribution of the radio pulsars. In a model for the evolution of the magnetic field, where the magnetic field is not confined to the crust such as the model of Konenkov and Geppert \cite{Gep2}, the relation between the equation of state and the magnetic filed decay is apparent as well. Here we did not consider the effect of various equations of the states. However, we propose that a theoretical model such as this model can be compared with the observational $P_s-B_s$ distribution of pulsars to rule out some of equations of the states. In fact, the numerical result of this paper can be extended to various equations of the states and cooling scenarios. 
The population synthesis is a necessary calculation for setting a limit on possible equations of states within the framework of the present model. These calculations can be done in future works. However, the main goal of this paper is to remind a reader of the importance of the extension of the results of \cite{Gep2} for single neutron stars to binary neutron stars and to present the result of such an extension in a few examples. 
In this paper we investigate the history of recycled binary neutron stars taking into account the various effects associated with the superconductive, superfluid core. We show that the presence of these effects modifies the evolution of the surface magnetic field substantially. The surface magnetic field of a binary neutron star with superfluid, superconductive type II core may increase during the accretion phase. We ask whether this alters the formation of MSPs. We show that despite the increase of the core magnetic field, MSPs are formed. In Sec. II we present a review of the standard evolution of binary neutron stars. The model for the evolution of the surface and core magnetic field  is explained in Sec. III. In Sec. IV we present our model for the evolution of binary neutron stars combining the general features of the Sec. II and Sec. III. In Sec V we give an overview of the corresponding numerical results.
Throughout this paper we discuss the evolution of the orbital separation of a binary $a(t)$, the surface magnetic field of a neutron star $B_{s}(t)$, the core magnetic field of a neutron star $B_{c}(t)$, the mass of a neutron star $M_n(t)$, and the mass of the companion $M_2(t)$. Therefore, any quantity depending on the above quantities also depends on time implicitly. Throughout this paper we avoid repeating the dependence on time of various quantities. However, the reader should be aware of this dependence unless otherwise stated.

 The main features of the model such as: 1) the increase of the strength of the magnetic field in the accretion phase, 2) the independence of the final strength of the surface magnetic field from the initial values of the core or surface magnetic fields, 3) the crucial role of the minimal spin period of the pulsar on the magnetic field evolution, 
are independent of the chosen parameters or the equation of state.
 \section{Evolution of neutron stars in binaries}
The weak magnetic field and fast rotational spin of MSPs suggest the idea of enhanced magnetic field decay in accreting neutron stars. According to the
standard picture, MSPs are old neutron stars spun up
by accretion of matter from their companion in binary systems. The accretion also heats up the crust of a neutron star. Hence, accretion enhances the decay of the magnetic field due to the decrease of the crustal conductivity. Therefore, in order to track the formation of MSPs in binaries, 
 we consider a model for orbital, spin, and magnetic field evolution of a newly born neutron star of mass $M_n$ and radius $R_n$ in a binary with initial orbital period $P_{oi}$. The companion has mass $M_2$ and radius $R_2$. 
In the study of binaries one might consider different companions.  Table I presents the values of different parameters that can be adopted for different companions. In the case of a low mass companion, wind accretion does not produce X-ray emission. However, the long main sequence lifetime of the companion allows for a long evolution in different phases. 
 The companion loses mass in the form of a spherical uniform stellar wind at rate $\dot{M_{2}}$. The orbital separation of the binary and consequently orbital period, change during the evolution due to the accretion, mass loss from the binary, and two other sinks of orbital angular momentum, namely, magnetic braking and gravitational radiation. In the case of a low mass companion, the orbital separation of the binary remains almost constant during the stellar wind evolution. However, for short initial orbital periods orbital evolution can alter the evolution of the surface magnetic field dramatically at the end stages of the evolution. Fig \ref{Ba} shows that including the effect of orbital evolution can change the final value of the final surface magnetic field $B_{sf}$ up to one order of magnitude.
Therefore, we do take into account the effect of the loss of the angular momentum from the system.
 \begin{table}
\caption{\label{tab:table2}Values of mass $M_2$, lifetime $\tau_2$, mass-loss rate $\dot{M}_2$ and wind velocity $V_w$ for low-mass (LM), intermidiate-mass (IM), and high-mass (HM), and Be-type companion stars (CS). }
\begin{ruledtabular}
\begin{tabular}{ccccccc}
$CS$ & $M_2$ & $\tau_2$ & $-\log\dot{M}_2$ & $V_w$\\
$~$ & $\mathrm{(M}_\odot{\mathrm)}$ &$\mathrm{(yr)}$ & $\mathrm{(M}_\odot \mathrm {yr}^{-1})$ & $\mathrm{ (km s}^{-1})$\\ \\
\hline\ \\
$LM$ & $0.8-1.0$ & $3\times10^{9}-10^{10}$ & $12-16$ & $400-800$\\
$IM$ & $2.5-7$ & $5\times 10^{7}-7\times 10^8$ & $9-11$ & $500$\\
$HM$ & $>9$ & $1.2\times 10^7-3\times 10^7$ & $6-10$ & $600-700$\\
$Be$ & $9$ & $3\times 10^7$ & $ 8-10$ & $200$
\end{tabular}
\end{ruledtabular}
\end{table}
 
 In a spin-induced field expulsion scenario (SIF) \cite{Ja2},
the evolution of the spin period and the magnetic field of a neutron star would be coupled. Our model is different from the original SIF model, as it will be explained later, but the coupling feature still persist in this model.

 \subsection{The binary neutron star vs. phases}
 The
standard evolution of neutron stars in close binaries is described in four
successive phases: the ejector, the propeller, the wind accretion, and the Roche Lobe overflow \cite{Ur}. In principle these phases might not be considered successively. A neutron star might return to a previous phase. However, the conditions for entering a phase can be different from that of returning back to a phase. This also brings a lot of complications on the temperature of the neutron star versus time. For example, for the transitions between the propeller and the accretion phase, a fight between the free cooling of the neutron star and heating due to the accretion arises. These conditions  have to be well studied before being included in any numerical computations. One should be aware that including such possibilities will change our picture of the evolution drastically. For example, a neutron star might reach almost a constant value of  magnetic field and spin period (before reaching an appropriate value for a MSP) while oscillating between accretion and propeller phase. This might alter the formation of the MSPs. Here, We consider that these phases occur successively. Therefore, all our results and conclusions depend on this assumption in the first place. Here, we give a brief overview of these phases.

{\em Ejector:}
A neutron star strong radiation pressure keeps the companion stellar wind matter
away from its magnetosphere. Hence, a neutron star does not experience
any influence from its companion and evolves like  an isolated
neutron star (single neutron star). During this phase  the magnetic field of a neutron star decreases and its spin period increases. We assume that during this phase a neutron star cools down according to the standard cooling scenario \cite{Van}.

{\em Propeller:} Due to the decrease of magnetic field the pulsar radiation pressure decreases.
Consequently, at some stage the neutron star radiation pressure is not sufficient enough to  keep the companion stellar wind matter
away from its magnetosphere, and
the stellar wind penetrates into the neutron star magnetosphere.
The magnetosphere of the neutron star
ejects the infalling wind matter transferring some part of the neutron star angular
momentum to the wind. Therefore, the neutron star spins down. The accretion torque produced by the capture of the plasma from the stellar wind depends on the detailed properties of the outflow.
A steady Keplerian disk may form from the accretion flow outside the magnetosphere. 
Spherically symmetric radial infall may also occur for these systems \cite{Hu}-\cite{Wa}. We use a coefficient $\zeta$ to count for the details of the flow (\ref{eq2-6}). In a more accurate model, numerical simulations of the flow should be included as well.
  In
this phase thermal and magnetic field evolution of neutron
stars do not differ from that of single neutron stars.

{\em Wind accretion:} Finally, the magnetic field
of a neutron star decreases sufficiently enough that accreted matter falls
onto its surface. In this phase the thermal, magnetic field, and spin evolution of a neutron star
differs completely from that of a single neutron star. Accretion with an accretion rate $\dot{M}_2=10^{-10}-10^{-9}M_{\odot}$yr$^{-1}$ or higher heats up the neutron star to an accretion temperature $T_{ac}=(2-3)\times 10^{8}$K \cite{Ha} and \cite{Fu}. 

{\em The Roche lobe overflow:} This phase begins when the
companion star fills its Roche lobe, so the
mass transfer is strongly enhanced. Accretion resulting from Roche lobe can last as long as $(1-5)\times10^7$yrs in low mass binaries. We assume that a neutron star enters Roche lobe overflow when either the companion leaves its main sequence (for the life time of different mass stars refer to Table I) or 
the radius of the companion $R_2$ equals to the Roche lobe radius $R_L$. This is possible to happen because we are taking into account the loss of orbital angular momentum of the binary and decrease in the binary radius $a$; however this is very unlikely for a low main sequence companion \cite{Note1}.
A measure of the Roche lobe radius is given by  the radius $R_L$ of a sphere which has the same volume as the lobe (the critical equi-potential surface which passes through the inner Lagrangian point $L_1$). A simple expression for the Roche lobe radius $R_R$ is given only in term of the ratio of the mass of the two stars $q=M_2/M_n$ \cite{Pa},
\begin{eqnarray}
R_R&=&a(0.38+0.2 \log q)~~~~(0.5<q<20)\nonumber\\
R_R&=&a(0.462\left(\frac{q}{q+1}\right)^{\frac{1}{3}})~~~(0<q<0.5),
\end{eqnarray}
where $a$ is the orbital radius of a binary \eq{Or1}.

We shall perform our numerical calculations only for a low main sequence companion. The computations are performed for a lifetime equal to the expected main-sequence lifetime of the companion star. Computations are stopped before the Roche-lobe overflow. Note that there is no difficulty in extending the results for other companions and beyond Roche-lobe.

To define the phase of a binary neutron star, we compare the accretion radius $R_a$, the radius of magnetosphere $R_m$, and the corotation radius $R_{co}$. Let us first define each of these  radii. 

 Stellar wind matter of impact parameter less than $R_a$, the so-called accretion radius, is absorbed by a neutron star gravitational field.  The accretion radius is given by
\begin{eqnarray}
R_a=\frac{2GM_n}{(V_w^2+V_o^2)},\label{equation acc}
\end{eqnarray}
where
\begin{equation}
V_{o}=\frac{M_2\sqrt{G}}{\sqrt{a(M_n+M_2)}},\label{vorb}
\end{equation}
is the orbital velocity of a neutron star and $V_w$ is the velocity of the wind
from its companion \cite{Ho}. The wind velocity changes when the distance from the star increases. The wind is believed to be accelerated to a typical terminal velocity of a few times the escape velocity at the stellar surface \cite{Ab}. However, we consider a constant wind velocity  assuming that the wind velocity close to a neutron star has already achieved its terminal velocity. Therefore, $V_w$ is considered as a constant parameter in our model.

Not all the mass lost from companion in the form of  stellar wind is accreted by a neutron star. The accretion rate reads 
\begin{eqnarray}
\dot M_{a}=\frac{\dot{M}_2 G^2 M_n^2}{r^2({V_{rel}^2+C_\infty^2})^{\frac{3}{2}}},
\end{eqnarray}
where $C_\infty $ is the speed of sound in the stellar wind matter, $V_{rel}=\sqrt{V_w^2+V_o^2}$, and  $r$ is the distance from the companion \cite{Da}. We assume that $\dot{M}_2$ is constant during the life-time period of the companion. 
The energy radiated by a rotating neutron star in  the form of relativistic particles or electromagnetic radiation originates at the characteristic distance of the so-called light cylinder radius
\be
R_l=\frac{c}{\Omega_s},
\ee
where $c$ is the speed of light and $\Omega_s=2\pi/P_s$ is the spin angular velocity of a neutron star.
The stellar wind matter is a highly conductive plasma, and at distances $r<R_l$ the electromagnetic field is static. Therefore, the wind can not penetrate through the field, but it can squeeze the field. The radius of the magnetosphere $R_m$ is defined where
the  pressure of the magnetic field around a neutron star and the pressure of the matter balance each other.
At a radius $r>R_a$ the pressure of the stellar wind $P_{w}(r,t)$, is given by the following equation
\begin{equation}
P_{w}=\frac{\dot M_2V_w}{4\pi(a-r)^2},\label{eq2-1}.
\end{equation}
 At a radius $r<R_a$ the stellar wind pressure grows as $r^{-\frac{5}{2}}$, and it is given by
\begin{equation}
P_{w}=r^{-\frac{5}{2}}\dot M_{a}\sqrt{\frac{GM_n}{8\pi^2}}.\label{eq2-2}
\end{equation}
If at $r=R_a$ the pressure of the dipolar magnetic field of a neutron star is smaller than the pressure of the stellar wind matter, the radius of magnetosphere reads
\begin{equation}
R_m=\left(2GM_n\dot M_{a}^2\right)^{-\frac{1}{7}}\left(B_sR_n^3\right)^{\frac{4}{7}},\label{eq2-3}
\end{equation}
where $B_s$ is the strength of the magnetic field on the surface of the neutron star \cite{Ja1}. For the case of a disk accretion, $R_m$ is similar. However, both smaller and larger values have been given by different authors \cite{Na} and \cite{Wa}.
If the radius of magnetosphere is larger than $R_a$, it can be found from
the equation
\begin{equation}
\left(\frac{2 |\dot M_2|}{B_s^2R_n^6}\frac{V_{rel}^2}{V_w}\right)^{\frac{1}{2}}R_m^3+\left(R_m-a\right)=0.\label{eq2-4}
\end{equation}
The magnetosphere of a neutron star is bounded by the light cylinder of radius $R_l$ \cite{Ja1}.

 To find the condition for the transition of a pulsar from the ejector to the propeller phase we first introduce the radiation pressure $P_{rad}$, pressure due to particles and radiation field emitted  by a neutron star at the distance $r$ from the star,
 \be
 P_{rad}=\frac{L_{rad}}{4\pi r^2 c},
 \ee
 where $L_{rad}$ is the total rate of the energy loss by the neutron star
 \be
 L_{rad}=\frac{2}{3c^3}R_n^6B_s^2  \Omega_s^4.
 \ee
 $R_{rad}$ is defined as the radius where the radiation pressure 
and the pressure of the stellar wind matter balance each other. 
Here we have two cases:

(1) $R_l<R_{rad}$ ($P_{rad}>P_{gas}$ at $r=R_l$): a neutron star will be in the ejector phase or the (most probably obscured) active radio pulsar phase.

(2) $R_{rad}>R_l$ ($P_{rad}<P_{gas}$ at $r=R_l$): the accreted matter makes contact with the magnetosphere and may cause the spin-up or spin-down of a neutron star.
When the matter reaches 
the magnetosphere it interacts with the neutron star; hence the angular momentum of the neutron star changes by the following equation
\begin{equation}
\dot L_s=-\zeta\dot M_{a}R_mV_{d}(r=R_m),\label{eq2-6}
\end{equation}
where $\zeta$ is an efficiency factor included to take into account the details of the geometry of the interaction ($\zeta=1$ corresponds to the case of disk accretion), and $V_{d}(R_m)=V_{co}(R_m)-V_{kep}(R_m)$ is the difference between the corotation velocity of stellar wind matter, $V_{co}(R_m)=\Omega_sR_m$, and Keplerian speed of the stellar wind matter around the neutron star, $V_{kep}(R_m)=\sqrt{\frac{GM_n}{R_m}}$, both evaluated at distance $R_m$ from the neutron star.

Define the radius of the corotation surface in the following form 
\begin{equation}
R_{co}=\left(\frac{GM_n }{\Omega_s^2}\right)^{\frac{1}{3}}.\label{eq2-5}
\end{equation}
When $R_{co}<R_m$, a neutron star is in propeller phase and the spin period of the neutron star decreases, but when $R_{co}>R_m$ the star is in wind accretion or spin-up phase. The amount of spin angular momentum added or extracted from the system is calculated assuming that: 1) In the ejector phase all the stellar wind of the companion leaves the system. 2) In the propeller phase all the accreted matter is expelled by the neutron star at its magnetosphere with an angular velocity equal to that of the neutron star. 
3) In accretion phase all the accreted matter enters the magnetosphere. In all these cases the fraction of the stellar wind that does not accrete to the neutron star leaves the binary with the angular momentum equal to that of the companion.

The spin-up of a neutron star due to the accretion is limited to an equilibrium period given by \cite{Va}
\be
P_{eq}=2.4B_{s9}^{\frac{6}{7}}~R_{n6}^{\frac{16}{7}}~\left(\frac{M_\odot}{M_n}\right)^{\frac{5}{7}}\left(\frac{\dot{M}_{Edd}}{\dot{M}_a}\right)^{\frac{3}{7}}~~\mathrm{ms},\label{Peq}
\ee
where $B_{s9}$ is the surface magnetic field in the unit of $10^9$G, $R_{n6}$ is the radius of the neuron star in the unit of $10^{6}$cm, $\dot{M}_{Edd}$ is the maximum possible Eddington limit for the accretion rate given by
\be
\dot{M}_{Edd}=1.5\times 10^{-8}R_{n6}~~{\mathrm M}_{\odot} \mathrm{yr}^{-1}.
\ee
The spin-up line in the pulsar $B_s-P_s$ diagram is given by $P_{eq}(\dot{M}_{Edd})$, where $\dot{M}_a$ is set equal to $\dot{M}_{Edd}$. In our calculation we set this value as the minimum value of the period that a neutron star can get to due to the accretion. This limit plays a crucial role in our model.  This limit controls the increase of the core and the surface magnetic field of a neutron star in the accretion phase. 
\subsection{Orbital evolution}
We have a binary system composed of a neutron star of mass $M_n$ and its companion of mass $M_2$. 
We assume that both stars are in circular orbit around the center of mass. Then orbital angular momentum of the neutron star and its companion is given by $L_n=M_nR_n^2\Omega_o$ and $L_2=M_2R_2^2\Omega_o$, respectively. Here,
$\Omega_o$ is the orbital angular velocity of each star around the center of mass,
\be\n{Or1}
\Omega_o=\sqrt{\frac{GM}{a^3}}.
\ee
where
$a$ is the orbital separation (orbital radius) of the binary and $M=M_n+M_2$ is the total mass of the binary. 
For our binary system the orbital angular momentum , $L_o=L_n+L_2$, is given by
\be\n{Or}
L_o=M_nM_2\sqrt{\frac{Ga}{M}}.
\ee
Now, we assume that the companion is losing mass in the form of stellar wind. Some fraction of this mass will leave the system. Also, there are some other sources of the angular momentum loss from the system such as gravitational waves or magnetic braking. All these effects cause the dependence of $a,~M,~M_n,~M_2,$ and $L_{o}$ to time.
The change in $a$, affects the rate of change in the spin period of a neutron star $\dot P_s$. From \eq{Or} the orbital radius evolves according to the equation 
\begin{eqnarray}
\frac{\dot a}{a}&=&2\frac{\dot{L}_o}{L_o}+\frac{\dot{M}}{M}-2\frac{\dot{M}_n}{M_n}-2\frac{\dot{M}_2}{M_2}
,\label{eq2-7}
\end{eqnarray}
where \cite{Ja1}
\begin{equation}
\dot L_o=\dot{L}_{esc}+\dot L_{MB}+\dot L_{GW}.\label{eq2-8}
\end{equation}
Here,
\begin{equation}
\dot L_{MB}=-0.5\times10^{-28}f^{-2}k^2M_2~R_2^4~\Omega_o^3~~\mathrm{g~ cm}^{2}\mathrm{s}^{-2},\label{eq2-9}
\end{equation}
is due to magnetic braking \cite{Ver}, $\Omega_o$ is the orbital angular velocity of the binary, $f=1.78$, and $k=0.1$ \cite {Sm}.
The rate of the loss of the orbital angular momentum from the binary due to the gravitational radiation $L_{GB}$ is given by
\begin{equation}
\dot L_{GW}=-32\frac{G^{\frac{7}{2}}M_n^2M_2^2M^{\frac{1}{2}}}{5c^5a^{\frac{7}{2}}}.\label{eq2-10}
\end{equation}
Finally we have
\begin{equation}
\frac{\dot L_{esc}}{\dot M}=\beta\frac{L_2}{M_2},\label{eq2-11}
\end{equation}
where $L_{esc}$ is the rate of the loss of the angular momentum from the binary due only to the escape of  stellar wind matter \cite{Ja1} and $\beta$ is the fraction of
the angular momentum of the companion star that escapes from the binary when the rate of total mass loss from the binary is $\dot{M}=\dot{M}_a-\dot{M}_2$.
In the accretion phase $\beta$ is given by (see \cite{Sho} for details of the derivation)
\begin{equation}
\beta=1+\left(\frac{\alpha-1}{\alpha}\right)\left(\frac{R_mM^{\frac{3}{2}}|V_{d}|}{G^{\frac{1}{2}}a^{\frac{1}{2}}(t)M_n^2}\right),\label{eq2-12}
\end{equation}
and in the propeller phase by
\begin{eqnarray}
\beta&=&1+\left(\frac{\dot M_{a}}{\dot M}\right)\left[\left(\frac{M_2}{M_n}\right)^2+\left(\frac{R_m}{a}\right)^{\frac{1}{2}}\left(\frac{M}{M_n}\right)^{\frac{3}{2}}\right].\nonumber\\
\label{eq2-13}
\end{eqnarray}
Here, $\alpha=(1-\dot{M}_a/{\dot M})$ is the fraction of the stellar wind matter which leaves the binary without any interaction with the neutron star. The rest interacts with the magnetosphere and changes the rotational angular momentum of the neutron star. 
In terms of the above parameters, \eq{Or} can be written in the following form:
\begin{eqnarray}
\frac{\dot a}{a}&=&-2\biggl[1-(1-\alpha)\frac{M_2}{M_n}-\frac{\alpha}{2}\frac{M_2}{M}
-\alpha\beta\frac{M_n}{M}\biggl]\frac{\dot{M}_2}{M_2}\nonumber\\
&+&2\biggl[\frac{\dot{L}_{GB}+\dot{L}_{MB}}{L_o}\biggl].\label{eqA}
\end{eqnarray}

We assume that the initial values of the spin period and of the magnetic field of a neutron star born in a binary are independent of its orbital period.
\section{magnetic Field Evolution}

We assume that initially the magnetic field of a neutron star penetrates both into its core and crust. We do not discuss the mechanism of the creation of this magnetic field. We assume that in the crust the field is maintained by the currents. We restrict ourselves to a consideration of the decay of a field which is assumed to be dipolar outside the neutron star. 
The evolution of the magnetic field ${\bf B}$ in the crust of the neutron star is given by  \cite{Bha1} and \cite{Bha2}
\begin{equation}
\frac{\partial {\bf B}({\bf r},t)}{\partial t}=-\nabla \times\left(\frac{c^2}{4\pi\sigma}{\bf \nabla \times B}({\bf r},t)\right)
+{\bf \nabla}\times\left({\bf v \times B}({\bf r},t)\right),\label{eq1-1}
\end{equation}
 where 
\begin{equation}
{\bf v}=v{\bf e_r}=-\frac{\dot M_{a}}{4\pi r^2\rho}{\bf e_r},\label{eqvel}
\end{equation}
 is the velocity of the material movement, and $\sigma$ is the electrical conductivity in the crust. 
 The velocity of the material movement is zero in the ejector and propeller phases. The magnetic field $\bf{B}$ is obtained from a vector potential ${\bf B=\nabla\times A}$. Due to axial symmetry, in spherical coordinates, ${\bf A}=(0,0,A_\phi(r,\theta,\phi))$.
 Introducing
 \begin{equation}
  A_{\phi}(r,t)=\frac{S(r,t)\sin{\theta}}{r}=\frac{B_{s0}R_n^2s(r,t)\sin\theta}{r},
  \end{equation} 
  we have
 \begin{eqnarray}
B_{r}(r,t)&=&B_{si}R_n^2\frac{2s(r,t)}{r^2}\cos\theta,\nonumber\\
B_{\theta}(r,t)&=&-{B_{si}R_n^2}\frac{\sin\theta}{r}\frac{\partial s(r,t)}{\partial r},
\end{eqnarray}
where $B_{si}$ is equal to the value of the surface magnetic field at $\theta=\pi/2$, where the value of the surface magnetic field is maximum. $|{\bf B}|(r=R_c)=B_{si}\sqrt{3\cos\theta^2+1}$.
Therefore, from equation (\ref{eq1-1}) we get
\begin{equation}
\frac{\partial S(r,t)}{\partial t}=\frac{c^2}{4\pi\sigma}\left(\frac{\partial ^2 S(r,t)}{\partial r^2}-\frac{2S(r,t)}{r^2}\right)+v(r,t)\frac{\partial S(r,t)}{\partial r}.\\
\label{eq1-2}
\end{equation}
Boundary condition on the surface of a neutron star reads \cite{Gep}
\begin{equation}
\left.R_n\frac{\partial S(r,t)}{\partial r}\right\vert_{r=R_n}+S(R_n,t)=0.\label{eq1-3} 
\end{equation}
The above boundary condition ensures that the magnetic field is dipolar outside a neutron star.

 Conductivity comes mostly from the scattering of the electrons on phonons and on  impurities \cite{Ur2}.
The conductivity due to the scattering on phonons $\sigma_{ph}$ is found from
\begin{equation}
\sigma_{ph}=1.21\times10^{28}\frac{x^2}{T^2}\sqrt{T^2+0.084T_D^2}~~\mathrm{s}^{-1},\label{con1}
\end{equation}
where $T$ is the temprature which changes with time according to the standard cooling scenario (except in the accretion phase), and $T_D$ is the Debye temperature
\begin{equation}
T_D=2.4\times10^6x^{\frac{3}{2}}\sqrt{\frac{2Z}{A}}~~{\mathrm K},\label{con3}
\end{equation}
where $Z$ and $A$ are the atomic number and the mass number of the dominant element in the crust of a neutron star. For a table of the dominate elements at different densities in the BPS equation of state refer to \cite{Baym}.
Here,
\begin{equation}
x=\frac{P_F}{m_ec},\label{con2}
\end{equation}
where $P_F$ is the Fermi energy, $m_e$ is electron mass, $c$ is the speed of light.
$x$ can be written in terms of the density in the following  
 form
\begin{equation}
x=\frac{\hbar}{m_ec}\left(\frac{3\pi^2\rho Z}{Am_p}\right)^{\frac{1}{3}},
\end{equation}
where $m_e$ is the mass of electron, $m_p$ is the mass of proton, and $\hbar$ is reduced Plank constant.
The phonon conductivity decreases when $T$ increases, therefore additional heating due to the accretion accelerates the magnetic field decay. At very low temperatures $\sigma_{ph}$ is very large.
The conductivity due to impurities reads
\begin{equation}
\sigma_{imp}=8.53\times10^{21}x\Lambda_{imp}^{-1}\frac{Z}{Q}~~\mathrm{s}^{-1},\label{con4}
\end{equation}
where $\Lambda_{imp}$ is the Coulomb logarithm for impurity scattering (at $\rho>10^5$gcm$^{-3}$ one has $\Lambda_{imp}=2$), and Q is the parameter characterizing the concentration and charge of impurities, known as the impurity parameter \cite{Ur2}. Here we assume that $0.001\leqslant Q \leqslant 0.1$ and that $Q$ is constant throughout the crust for the whole period of the evolution. We assume that at very low temperatures $T<10^2$K, the conductivity is given by only the scattering on the phonons. A binary neutron star in most cases will not reach to such temperatures due to accretion (accretion starts before the cooling makes the neutron star so cold). \\
Finally, the total conductivity in the crust, $\sigma$, is given by
\begin{equation}
\frac{1}{\sigma}=\frac{1}{\sigma_{ph}}+\frac{1}{\sigma_{imp}}.\label{con5}
\end{equation}
Magnetic field evolution of the neutron star is sensitive to the temperature due to the dependence of the conductivity on the temperature. Accretion effects the crustal conductivity substantially. We assume that temperature is uniform throughout the crust. For more exploration of the current model one has to consider the dependence of the temperature to the density.

We assume that in the core of a neutron star protons form a superconductor of type II, and the magnetic flux is concentrated in the array of magnetic flux tubes called fluxoids.
The evolution of the magnetic field in the superconductive core of a neutron star is given by \cite{Gep2}
\begin{equation}
\frac{\partial {\bf B_c}({\bf r},t)}{\partial t}=\nabla \times \left({\bf v_p}({\bf r},t) \times {\bf B_c}({\bf r},t)\right)\label{eq1-4},
\end{equation}
where ${\bf v_p}(r,t)=v_p(r,t){\bf e_r}$ is the radial velocity of fluxoids.

The core of a neutron star consists of neutrons in a superfluid state and protons in a superconductor state. In steady state vortices in the core of a neutron star corotate with
the superfluid bulk matter at the rate $\Omega$. When a star spins down at a rate $\dot \Omega$ , the superfluid bulk matter rotates at a faster rate, $ \Omega_b$. We assume that if  $\dot \Omega$ is kept constant there is a constant difference (lag) between the angular velocity of the superfluid bulk matter and the neutron vortices $\omega_{l}=\Omega_b-\Omega_s>0$. The 
radial velocity of the vortices at the core-crust boundary is determined from the conservation of angular momentum \cite{Sir} and \cite{Ja2}
\begin{equation}
{\bf v}_n=-\frac{R_c{\dot\Omega_s}}{2\Omega_b}{\bf e}_r\cong-{R_c}\frac{\dot\Omega_s}{2\Omega_s}{\bf e}_r=R_c\frac{\dot{P_s}}{2P_s}{\bf e}_r,\label{eq1-5}
\end{equation}
where $R_c$ is the radius of the core. When a neutron star spins down $\dot{P_s}>0$, and vortices move outward, so one has ${\bf v}_n>0$. When the neutron star spins up, $\dot{P_s}<0$ and vortices move inward to the core, so one has ${\bf v}_n<0$. For a single neutron star $\dot{P_s}$ is always positive and vortices always move outward from the core. In a binary neutron star in the accretion phase, the neutron star spins up. Therefore, vortices start moving back into the core. 
In the original (SIF) model \cite{Ja2} the pinning between the neutron vortices and the proton fluxoids was suggested to result in the expulsion of the magnetic flux out of the core, into the crust of a neutron star at a rate equal to the spin-down rate of the star. However, due to the existing resistance against the motion of fluxoids and vortices,  fluxoids and vortices move with different velocities.
There are several forces acting on fluxoids driving them outward to the conductive crust of a neutron star, where we have an ohmic decay (\ref{eq1-2}).
 Therefore, to find the rate of the out-flow or in-flow of the magnetic flux we consider various forces which act on the fluxoids in the interior of a neutron star such as the force due to their pinning interaction with the moving neutron vortices, the viscous drag force due to the magnetic scattering of electrons, and the bouyancy force. The resulting motion of fluxoids in a binary neutron star is not trivial. However, there  exist a possibility for the outflow of the fluxoids.  
$B_c(R_c,t)$ gives us the appropriate inner boundary condition (at the bottom of the crust) for equation (\ref{eq1-2}) .
 The pinning per unit length exerted on a fluxoid is given by \cite{Mu}
\begin{eqnarray}
{\bf f}_n(R_c,t)=\frac{2\Phi_0\rho R_c\Omega_s(t)\omega_{l}(t)}{B_c(R_c,t)}{\bf e}_r~~\mathrm{dyn},\label{eq 1-10}
\end{eqnarray}
where  $\Phi_0=2\times 10^{-7}$G cm$^2$ is the quantum of the magnetic flux carried by the fluxoid, $\rho$ is the density in the core-crust boundary of a neutron star, and
$B_c(R_c,t)$ is the strength of the magnetic field in the core-crust boundary in units of G \cite{Din}. Pinning can be negative or positive.
The maximum $\omega_{l}$ which can be sustained by the pinning force defines the maximum force which can be exerted on fluxoids by vortices.
This maximum value of $\omega_{l}$ is known as $\omega_{cr}$ \cite{Din}
\begin{eqnarray}
\omega_{cr}&=&1.59\times 10^{-6} B_{c8}^{\frac{1}{2}}~\mathrm{rad}~\mathrm{s}^{-1},
\label{eq1-14}
\end{eqnarray}
where $B_{c8}$ is the strength of the core magnetic field in the units of $10^{8}$G.\\
The bouyancy force per unit length exerted on  fluxiod reads \cite{Mu}
\begin{eqnarray}
{\bf f}_b=\frac{1}{R_c}\left(\frac{\Phi_0}{4\pi \lambda_p}\right)^2\ln\left(\frac{\lambda_p}{\xi}\right){\bf e}_r~~\mathrm{dyn}.\label{eq1-12}
\end{eqnarray}
Note that
\begin{equation}
\lambda_p=1.315\times 10^{-15}~~\mathrm{cm},
\end{equation}
$\lambda_p$ is the London penetration depth, where $\frac{\lambda_p}{\xi}>\sqrt{2}$ for a superconductor type II.
The bouyancy force is always positive, that is it is always directed outward.
The drag force per unit length exerted on fluxoid is given by
\begin{eqnarray}
{\bf f}_v(R_c,t)=-\frac{3\pi}{64}\frac{n_e~e^2\Phi_0^2}{E_F\lambda_p}\frac{{\bf v}_p(R_c,t)}{c} ~~\mathrm{dyn},\label{eq1-13}
\end{eqnarray}
where $n_e$ is the number density of electrons, $E_F$ is the electron Fermi energy. Here we use the values $n_e=3\times 10^{36}$ cm$^{-3}$,
$n_n=1.7\times 10^{38}$ cm$^{-3}$, and $E_F=88$ Mev \cite{Jo}. When ${\bf v}_p>0$, that is when fluxoids are moving outward ${\bf f}_v<0$.

It was shown in \cite{Gol} that the migration of fluxoids depends on the rate of the URCA process due to the fact that it is dependent on the beta equilibrium and the rate it is restored. Here we consider the model proposed by \cite{Jo}, since we are closely following the formalism developed by \cite{Gep2} and \cite{jaa}. We are interested in comparing our results with \cite{Gep2}. Namely, we want to compare the single and binary stars while we keep the formalism of the magnetic field evolution the same. One may improve the model considering the fluxoids drag derived in \cite{Gol} rather than \cite{Jo}.
\\
Assuming that the magnetic flux in the core is uniform, we can find the total forces exerted on fluxoids \cite{Gep2}
\begin{eqnarray}
{\bf F}_{n,b,v}={\bf f}_{n,b,v}\frac{4R_c}{3}N_p,\label{eq1-15}
\end{eqnarray}
where $4/3R_c$ is the mean length of the fluxoids, and $N_p=\pi R_c^2B_c/\Phi_0$ is the number of fluxoids.
The net power due to the forces is equal to the Poynting flux through the surface of a neutron star core
\begin{equation}
\sum_{fluxoids}\int(f_b+f_n+f_v)v_p ~dl=
-\frac{c}{4\pi}\int_{Score}[{\bf E\times B}].{\bf da}_{core},\label{eq1-16}
\end{equation}
where the integration on the right hand side is done over the surface of the core, the normal vector of this surface is being directed inward \cite{Gep2}.
This equation can also be written in the form \cite{Gep2}
\begin{eqnarray}
{\bf F}_n+{\bf F}_b+{\bf F}_v({\bf v}_p)+{\bf F}_{c}=0.\label{eq1-17}
\end{eqnarray}
The above equation gives us the radial velocity of the fluxoids, $v_p$.
From equation (\ref{eq1-16}) one can find the appropriate form of $F_{crust}$.
We have
\begin{eqnarray}
{\bf\nabla} \times {\bf B}=\frac{4\pi}{c}{\bf J},\label{eq1-18}
\end{eqnarray}
therefore, we find
\begin{eqnarray}
{\bf E}&=&\frac{c}{4\pi \sigma}\left(\frac{\sin\theta}{r}\right)\left(-\frac{\partial^2S}{\partial r^2}+\frac{2S}{r^2}\right){\bf e}_\phi .\nonumber\\
\label{eq1-19}
\end{eqnarray}
and using (\ref{eq1-1}) we find
\begin{eqnarray}
{\bf F}_{c}(R_c,t)&=&\frac{2}{3}\frac{1}{v_p}~\left.\left(\frac{\partial S(r,t)}{\partial t}\right)\right\vert_{r=R_c}\left.\left(\frac{\partial S(r,t)}{\partial r}\right)\right\vert_{r=R_c}\nonumber\\
&-&\frac{c^2v_c}{6\pi\sigma_c}\left(\frac{\partial S(r,t)}{\partial r}\right)^2\vert_{r=R_c} {\bf e}_r, 
\label{eq1-20}
\end{eqnarray}
where $\sigma_c$ is the conductivity at the core-crust boundary and $v_c=\dot{M}_{a}/(4\pi R_c^2\rho_c)$. The second term is $10^{20}$ times smaller than the first term, and is zero except in the accretion phase. We neglect this term.
Making the same assumptions as we have made deriving equation (\ref{eq1-2}), we can write equation (\ref{eq1-4}) in the from
\begin{equation}
\frac{\partial S(r,t)}{\partial t}=-v_p(r,t)\frac{\partial S(r,t)}{\partial r}.\label{eq1-10}
\end{equation}
We assume that the core magnetic field is homogeneous, therefore \cite{Gep2}
\begin{equation}
S(r,t)=\frac{B_c(t) r^2}{2}.\label{eq1-11}
\end{equation}
Equations (\ref{eq1-11}) and (\ref{eq1-10}) give
\begin{equation}
v_p(r,t)=\gamma(t)r,\label{eq1-12}
\end{equation}
Making the assumptions above, we have
\begin{equation}
S(r=R_c,t)=S(r=R_c,t_0)\exp({-2\int_0^t \gamma(t^{'}) dt^{'}}),\label{eq 1-13}
\end{equation}
where $S(R_c,t_0)$ is the value of the stream function in the core-crust boundary at the initial time $t_0$. When fuxoids are transported out of the superconductive core (${\bf v_p}>0$), the value of the stream function $S$ at the core-crust boundary decreases. When ${\bf v_p}<0,$ fluxoids move back into the core and the magnetic field in the core increases. (\ref{eq1-13}) gives us the inner boundary condition \cite{Gep2} for equation (\ref{eq1-2}), $S(R_c,t)$.
The increase of $S(R_c,t)$ (when ${\bf v}_p>0$), leads to the increase of the surface magnetic field of a neutron star.
From (\ref{eq1-11} and \ref{eq1-13}) $F_c$ is given by (the second term is neglected)
\begin{eqnarray}
{\bf F_c}=-\frac{2}{3}B_c^2R_c^2~{\bf e}_r
\end{eqnarray}
Therefore, ${\bf F}_c$ is always negative.
\section{Description of the model }

We assume that $t>10^8$ yrs of the evolution of binary neutron stars  will result in the formation of MSPs. Accretion of matter from the stellar wind of the companion affects the magnetic field evolution due to the change of the crustal conductivity, the orbital evolution, and most importantly the spin evolution of a neutron star.
Here, we consider the coupled spin, mass, magnetic field, orbital separation, and orbital period evolution of a neutron star in a binary. Namely we combine the magnetic field model discussed in Sec. III to the standard spin, and orbital evolution of a neutron star. We also take into account the cooling of the neutron star. Note that the magnetic field model we consider here was previously considered only for single neutron stars.  

In the ejector, propeller and accretion phase we integrate a set of the differential equations (to be explained later), to find the evolution of quantities $M_n(t)$, $a(t)$, $P_s(t)$, $s(t)$, $B_s(t)$ and $B_c(t)$. To integrate equation (\ref{eq1-2}) we have to know the initial stream function and the inner (\ref{eq1-13}) and the outer boundary conditions. $\gamma$ in (\ref{eq 1-13}) is proportional to the velocity of the fluxoids. The velocity of the fluxoids, ${\bf v}_p$, is given by equation (\ref{eq1-17}). However, in this equation $\omega_{l}$ is unknown as well (since $F_n$ depends on $\omega_{l}$). The following strategy helps us on the derivation of ${\bf v}_p$.
Vortices can move faster than fluxoids if $\omega_{l}=\omega_{cr}$, or they can move with the same velocity if $-\omega_{cr}<\omega_{l}<\omega_{cr}$, or they move slower than fluxoids if $\omega_{l}=-\omega_{cr}$.
To perform the numerical calculations, we start with $v_p(R_c,t_0)=v_n$. This is a preliminary assumption, and we use it to calculate $F_b$, $F_v$, $F_{crust}$, and $F_n$ to find $\omega_{l}$ from equation (\ref{eq 1-10}). Then, to find the actual velocity of the fluxoids, we check the following conditions.
If $-\omega_{cr}<\omega_{l}<\omega_{cr}$,  fluxoids and vortices comove, and we set $v_p=v_n$.
If $\omega_{l}>\omega_{cr}$, the result derived from the preliminary assumption is not correct. Since, $\omega_l$ can not be greater than $\omega_c$. Therefore, we set $\omega_{l}=\omega_{cr}$, and calculate $F_n(\omega_{cr})$, where $v_p$ is given by equation (\ref{eq1-17}). In this case vortices move faster than fluxoids.
If $\omega_{l}<-\omega_{cr}$,  we  set $\omega_{l}=-\omega_{cr}$, and calculate $F_n(-\omega_{cr})$. Equation (\ref{eq1-17}) gives us $v_p$, and therefore $\gamma$ (see equation \ref{eq1-12}). In this case vortices move slower than fluxoids.
Having $v_p$ , we calculate $S(R_c,t)$ from equation
(\ref{eq 1-13}). This gives us the inner boundary condition for equation (\ref{eq1-2}).
In our numerical calculations we use a normalized stream function $s(r,t)=S(r,t)/(B_{s0}R_n^2)$, where $B_{s0}$ is the initial value of surface magnetic field. The normalized stream function $s$ has no units. Surface magnetic field in terms of the normalized stream function $s$ is given by $B_s(t)=B_{s0}s(R_n,t)$. Therefore, from (\ref{eq1-11}) the core magnetic field is given by
\be
B_c(t)=2B_{s0}s(r,t)\left(\frac{R_n}{R_c}\right)^2.
\ee
We assume that the initial profile of the normalized stream function is given by
\be
s(r,t_0)=1-s_c\left(\frac{R_n-r}{R_n-R_c}\right)^2,
\ee
where $0\leqslant s_c\leqslant 1$. $s_c=1$ gives an initial profile with no magnetic field in the core, while $s_c=0$ gives a profile of uniform magnetic field. In terms of this initial profile the initial ratio of core to the surface magnetic field is given by
\be
\frac{B_{c0}}{B_{s0}}=2(1-s_c)\left(\frac{R_n}{R_c}\right)^2.
\ee
It is possible to consider that initially $B_{s0}=B_{c0}$. However, this is not the only possibility, and early thermodynamic instabilities of the crustal layers might lead to $B_{s0}>B_{c0}$. 
\begin{table}
\caption{\label{tab:table 2} Some trial values of the initial magnetic field of the neutron star, $B_{si}$, intitial rotational (spin) period
of the neutron star, $P_{si}$, the initial orbital period of  the binary, $P_{oi}$, the velocity of stellar wind, $V_{w}$, the rate of mass loss from a low main sequence companion, $\dot M_2$, impurity parameter, $Q$, and the efficienty factor, $\zeta$ is shown in the Table. }
\begin{ruledtabular}
\begin{tabular}{ccccccc}
$B_{si}$ & $V_w$ & $P_{si}$ & $Log(\dot M_2)$ & Q & $M_2$ & $\zeta$\\
$\mathrm{(G)}$ & $\mathrm{(km s}^{-1})$ & $\mathrm{(s)}$ & $\mathrm{(M}_\odot~\mathrm{yr}^{-1})$ & & $\mathrm{(M}_\odot)$ & \\ \\
\hline
\\
$10^{14}$& 800 &1& -13&1&1&10\\
$3\times10^{13}$&600&0.4&-14&0.1&0.9&1\\
$10^{13}$&500&0.1&-15&0.01&0.8&0.6\\
$3\times10^{12}$&400&0.01&-16&....&....&0.1\\
$10^{12}$&....&....&....&....&....&0.02\\
\end{tabular}
\end{ruledtabular}
\end{table}
As it was mentioned before we consider the evolution of a neutron star in different phases.\\
(i) {\em Ejector phase}:  In this phase the orbital separation changes according to the equation (\ref{eq2-7}),
where all the matter which is lost from the companion leaves the binary, therefore
$\alpha=\beta=1$.
Since the stellar wind of the companion leaves the binary, the mass of the neutron star remains constant, $\frac{dM_n}{dt}=0$.
Due to the magnetodipole radiation the neutron star spins down at the following rate
\begin{eqnarray}
\frac{dP_s}{dt}=\frac{8}{3}\left(\frac{B_s^2R_n^6\sin^2 \chi}{c^3I}\right)\frac{1}{P_s},\label{eq3-8}
\end{eqnarray}
where $\chi$ the angle between the rotational and magnetic axis. We set this angle to $\pi/2$.
The crustal magnetic field evolution is governed by equation (\ref{eq1-2}), where the material movement is set to zero, $v=0$.
Equation (\ref{eq1-10}) gives the evolution of the stream function.

(ii){\em Propeller phase}:  In this phase we have the same set of equations, but instead of equation (\ref{eq3-8}) we use (\ref{eq2-6}). $V_{d}>0$ in this phase and the neutron star spins down.
In this case $\alpha=1$ and $\beta$ is given by equation (\ref{eq2-13}). Change in the mass of the neutron star is zero. In equation (\ref{eq1-2}), the velocity of the material movement in the crust is zero.

(iii) {\em Accretion phase}:  Accretion phase starts when $R_{co}>R_m$. In this phase some portion of the stellar wind is absorbed 
 by the neutron star, therefore $\alpha$ is equal to
\begin{eqnarray}
\alpha=1+\frac{\dot M_n}{\dot M_2},
\end{eqnarray}
where $\dot{M_n}=-\dot{M}_a$.
Finally, $\beta$ is given by equation (\ref{eq2-12}).
Accreted matter, when it reaches the magnetosphere, interacts with the neutron star. In this phase $V_{d}<0$, and the neutron star spins up
according to equation (\ref{eq2-6}).
The evolution of the magnetic field in the crust is governed by (\ref{eq1-2}), where $v$ is found from (\ref{eqvel}). 
 
\section{Numerical Results and Conclusion}
\begin{figure}[htb]
\begin{center}
\includegraphics[height=5.379cm,width=8.cm]{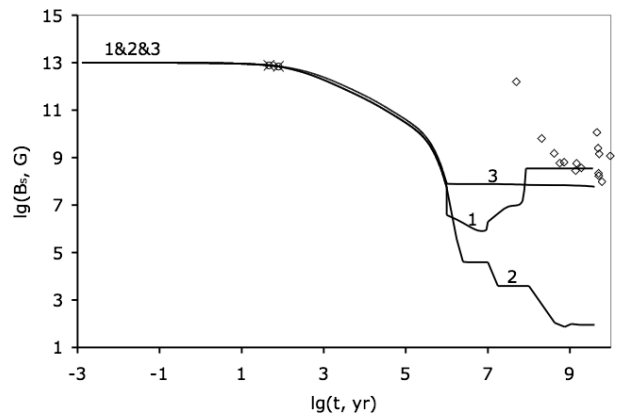} 
\caption{The evolution of the surface magnetic field for: (1) A binary neutron star with a low main sequence companion, the effect of the superconducting
core included. Initially the neutron star is in ejector phase. The first circle corresponds to the point of the transition to the propeller phase and the second circle corresponds to the point of the transition to the accretion phase. (2) A binary neutron star with the same companion, the effect of the superconducting core not included. Initially the neutron star is in ejector phase. The first cross corresponds to the point of the transition to the propeller phase and the second cross corresponds to the point of the transition to the accretion phase.
(3)A single neutron star, the superconducting core is included. The binary parameters are $M_2=$M$_\odot$, $R_2=$R$_\odot$, $P_{oi}=14$ days, $\dot{M}_2=10^{-12}$ M$_\odot/$yr, $\zeta=0.1$, $V_w=500$ km/s. Neutron star parameters are $P_{si}=0.01$ s,  $B_{si}=10^{13}$ G, $B_{ci}=10^{10}$ G, and $Q=0.01$. Diamonds are observed pulsars with low mass main sequence companions ($M_2=0.8-1.5$ M$_\odot$). At the given age the value of the surface magnetic field of these pulsars in known.}\label{Bt1}
\end{center}
\end{figure}
\begin{figure}[htb]
\begin{center}
\includegraphics[height=5.337cm,width=8.cm]{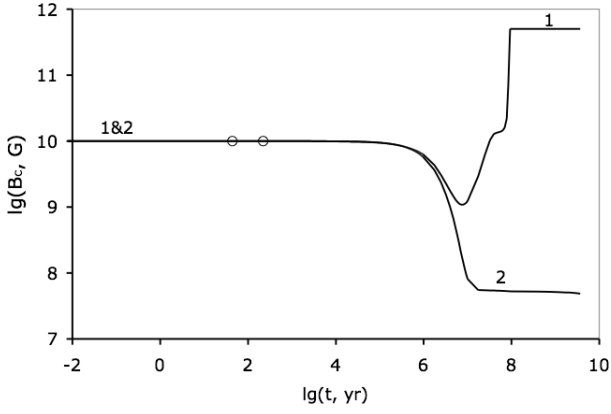} 
\caption{The evolution of the core magnetic field for: (1) the binary neutron star with a low main sequence companion,  (2) the single neutron star.
Parameters are the same as in Fig 1.}\label{Bc}
\end{center}
\end{figure}
\begin{figure}[htb]
\begin{center}
\includegraphics[height=5.329cm,width=8.5cm]{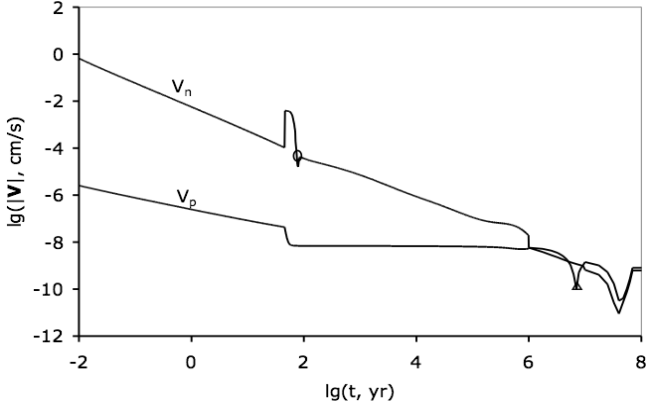} 
\caption{The evolution of the magnitude of the velocity of the neutron vortices $|{\bf V}_n|$ and proton fluxoids $|{\bf V}_p|$. The circle and  the triangle mark the points when the velocities change sign and become negative. Parameters are the same as in Fig 1.}\label{V1}
\end{center}
\end{figure}
\begin{figure}[htb]
\begin{center}
\includegraphics[height=5.337cm,width=8.cm]{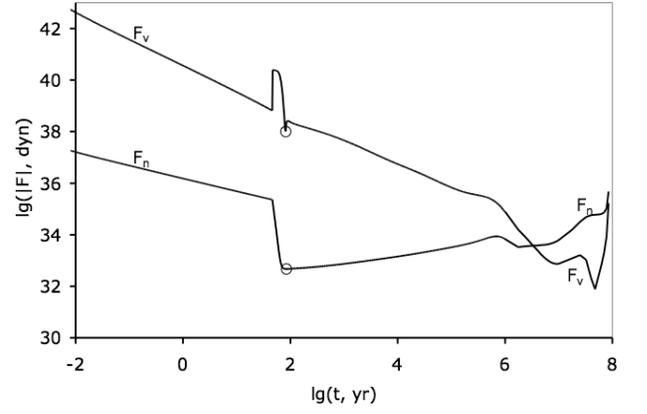} 
\caption{The evolution of the magnitude of the drag force $|{\bf F}_v|$ and pinning force $|{\bf F}_n|$. Circles mark the points when the forces change sign. ${\bf F}_n$ is negative after the circle and ${\bf F}_v$ is negative before the circle.
Parameters are the same as in Fig 1.}\label{F2}
\end{center}
\end{figure}
\begin{figure}[htb]
\begin{center}
\includegraphics[height=5.960cm,width=8.cm]{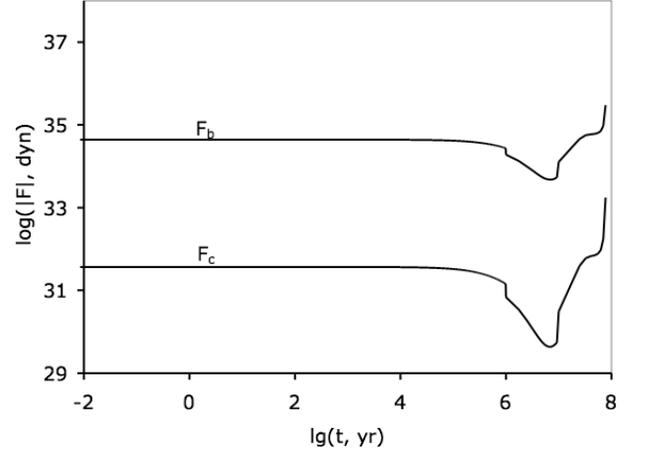} 
\caption{The evolution of the magnitude of the buoyancy force $|{\bf F}_b|$ and $|{\bf F}_c|$. The buoyancy force ${\bf F}_b$ is always positive and ${\bf F}_c$ is always negative.
Parameters are the same as in Fig 1.}\label{F1}
\end{center}
\end{figure}
\begin{figure}[htb]
\begin{center}
\includegraphics[height=5.370cm,width=8.cm]{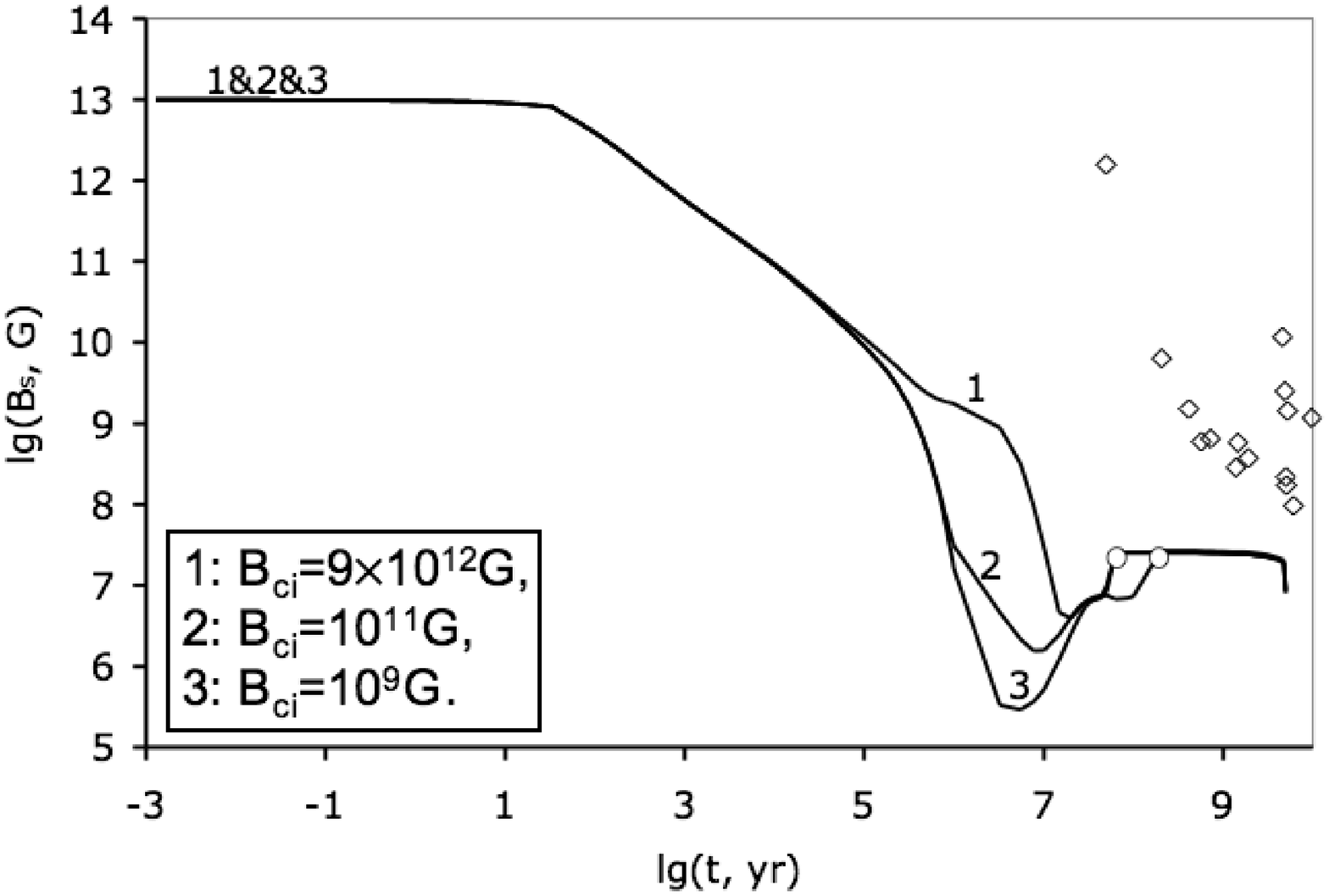} 
\caption{The evolution of the surface magnetic field for the binary neutron star corresponding to different values of initial core magnetic field $B_{ci}$.
Parameters are the same as in Fig 1.}\label{Bs1}
\end{center}
\end{figure}
\begin{figure}[htb]
\begin{center}
\includegraphics[height=5.370cm,width=8.cm]{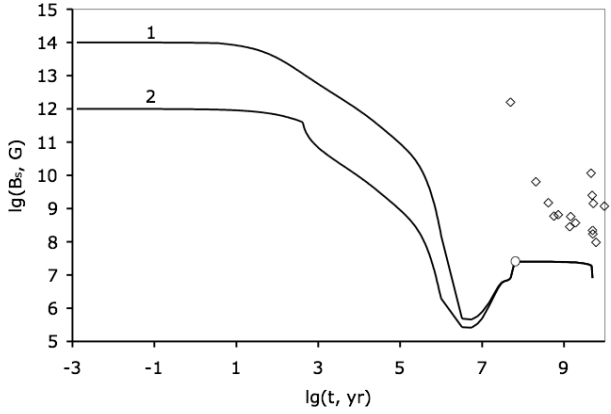} 
\caption{The evolution of the surface magnetic field for a binary neutron star corresponding to different values of initial surface magnetic fields $B_{si}$. (1) $B_{si}=10^{14}$ G. (2) $B_{si}=10^{9}$ G. Initial value of the core magnetic field is $B_{ci}=10^{9}$ G.
The rest of the parameters are the same as in Fig 1.}\label{Bs3}
\end{center}
\end{figure}
\begin{figure}[htb]
\begin{center}
\includegraphics[height=5.392cm,width=8cm]{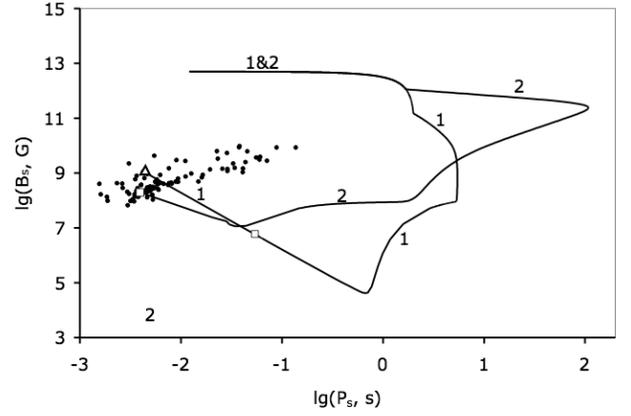} 
\caption{The evolutionary track of surface magnetic field $B_s$ and spin period $P_s$ for $t=4\times 10^9$ yrs. (1) $\dot{M}_2=10^{-14}$ M$_\odot/$yr. The square shows the end state of the evolution after $t=5\times 10^9$ yrs. The triangle shows the end state of the evolution after $t=7\times 10^9$ yrs.  (2)$\dot{M}_2=10^{-12}$ M$_\odot/$yr,
The square shows the end state of the evolution after $t=5\times 10^9$ yrs.
Parameters are: $P_{si}=0.01$ s, $B_{si}=5\times10^{12}$ G, $B_{ci}=10^{11}$ G, $P_{oi}=5$ days, $Q=0.3$, and $\zeta=0.1$. Black circles are observed millisecond pulsars.}\label{BP}
\end{center}
\end{figure}
\begin{figure}[htb]
\begin{center}
\includegraphics[height=5.342cm,width=8cm]{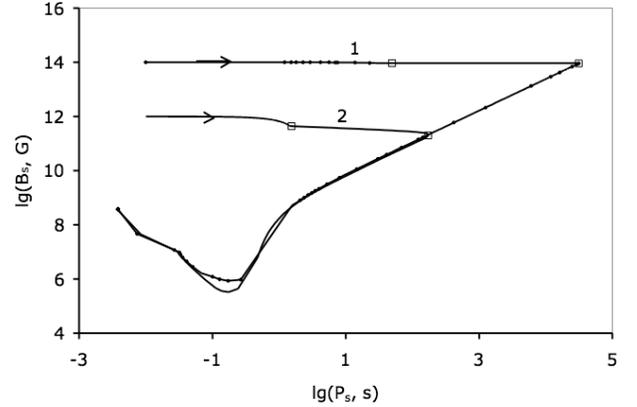} 
\caption{The evolutionary path of a binary neutron star in $B-P_s$ diagram corresponding to different values of initial surface magnetic fields $B_{si}$. (1) $B_{si}=10^{14}$ G. (2) $B_{si}=10^{10}$ G. Initial value of the core magnetic field is $B_{ci}=10^{10}$ G.
The rest of the parameters are the same as in Fig 1.}\label{BP3}
\end{center}
\end{figure}

\begin{figure}[htb]
\begin{center}
\includegraphics[height=5.404cm,width=8cm]{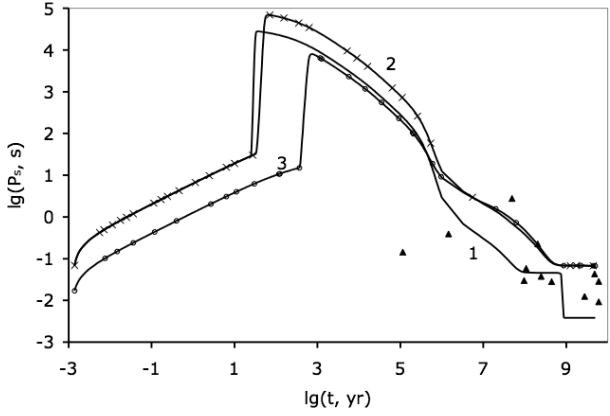} 
\caption{The evolution of the spin period of a binary neutron star. (1) $B_{si}=5\times10^{13}$ G and $\dot{M}_2=10^{-13}$ M$_\odot/$yr. (2) $B_{si}=5\times10^{13}$ G and $\dot{M}_2=10^{-14}$ M$_\odot/$yr. (3) $B_{si}=\times10^{13}$ G and $\dot{M}_2=10^{-13}$ M$_\odot/$yr. The initial value of the core magnetic field is $B_{ci}=10^10$ G and the initial orbital period is $P_{oi}=4$ days.
The rest of the parameters are the same as in Fig 1. Triangles are observed binary pulsars. These pulsars have an orbital period $0.1-3.5$ days at their current age. Their companions are low mass $M_2=0.8-1.3$ M$_\odot$. }\label{Pt}
\end{center}
\end{figure}
\begin{figure}[htb]
\begin{center}
\includegraphics[height=5.404cm,width=8.0cm]{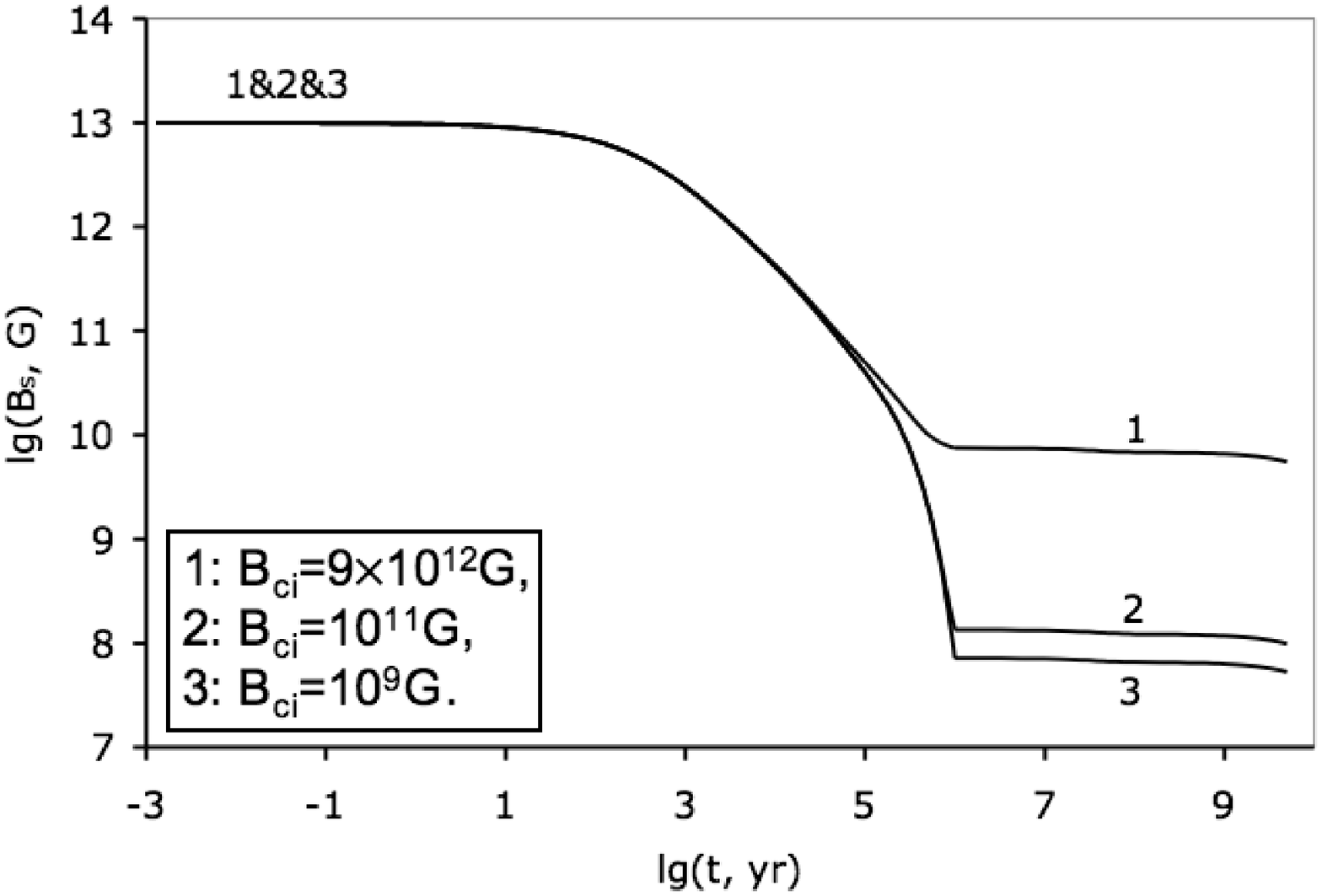} 
\caption{The evolution of the surface magnetic field for a single neutron star corresponding to different values  of initial core magnetic field $B_{ci}$.
Parameters are the same as in Fig 1.}\label{Bs2}
\end{center}
\end{figure}
\begin{figure}[htb]
\begin{center}
\includegraphics[height=5.283cm,width=8.5cm]{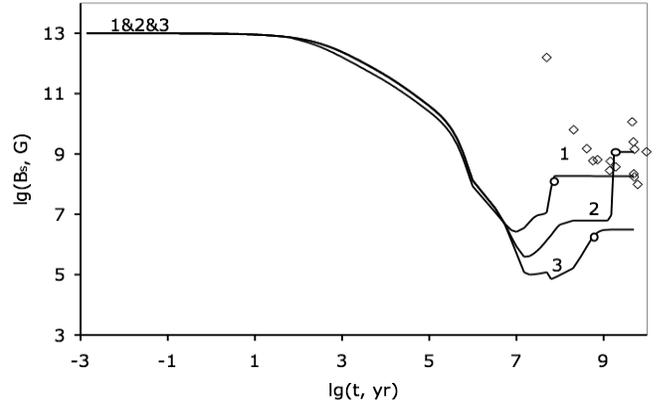} 
\caption{The evolution of the surface magnetic field for a binary neutron star corresponding to different values of $\dot{M}_2$. (1) $\dot{M}_2=10^{-12}$M$_\odot/$yr. (2) $\dot{M}_2=10^{-13}M_\odot/yr$. (3) $\dot{M}_2=10^{-14}M_\odot/yr$. Initial value of the core magnetic field is $B_{ci}=10^{11}G$, the initial value of the surface magnetic field is $B_{si}=10^{13}G$, the initial orbital period is $P_{oi}=10$ days, and $Q=0.01$.
The rest of the parameters are the same as in Fig 1. Circles mark the point of the evolution where $P_s=P_{eq}$ for the first time. Magnetic field increases by a tiny amount after this point.}\label{Bt5}
\end{center}
\end{figure}
\begin{figure}
\begin{center}
\includegraphics[height=5.366cm,width=8cm]{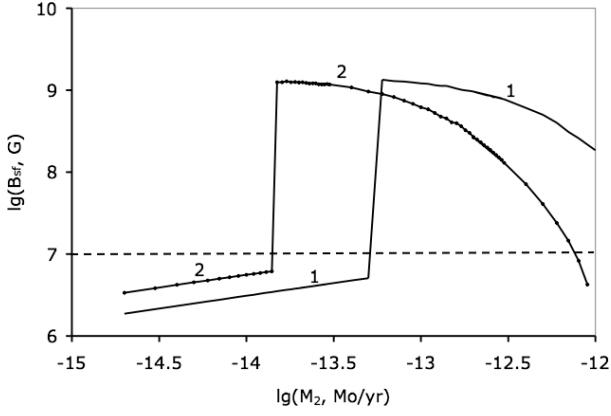} 
\caption{The dependence of the final surface magnetic field $B_{sf}$ on the rate of the mass loss from the companion $\dot{M}_2$. (1)$P_{oi}=4$ days (2)$P_{oi}=10$ days.
The initial strength of the surface magnetic fields is $B_{si}=10^{14}$ G, the initial strength of the core magnetic field is $B_{ci}=\times10^{9}$ G,  $Q=0.01$, $\zeta=0.1$, and the rest of the
parameters are the same as in Fig 1. The evolution has been considered for $5\times10^9$ yrs. Dash line marks the range of the strength of the surface magnetic field which is the most reasonable for millisecond pulsars $B_s=10^7-10^{10}$ G. The lowest value of the surface magnetic field observed is $6.67\times10^7$G which belongs to $J2229+2643$. This pulsar is $3.2\times10^{10}$ yrs old.}\label{BM2}
\end{center}
\end{figure}
\begin{figure}
\begin{center}
\includegraphics[height=6.110cm,width=8.0cm]{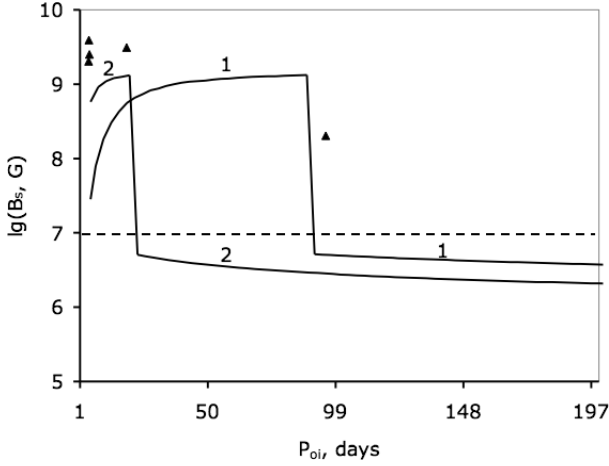} 
\caption{The dependence of the final surface magnetic field $B_{sf}$ on the initial orbital period of the binary $P_{oi}$ for two different values of the speed of the stellar wind: (1)$V_{w}=500$ km/s (2)$V_{w}=800$ km/s.
The initial strength of the surface magnetic fields is $B_{si}=10^{14}$ G, the initial strength of the core magnetic field is $B_{ci}=\times10^{12}$ G,  $Q=0.01$, $\zeta=0.1$, and the rest of the
parameters are the same as in Fig 1. The evolution has been considered for $5\times10^9$ yrs. Triangles are observed pulsars. These pulsar have an age $8.78\times10^{8}-4.83\times10^{9}$ yrs, and low mass companions $M_2=0.8-1M_{\odot}$. Their initial orbital period have been approximated based on their current orbital period within our code. Squares mark the point of the transition from the ejector to the propeller and from the propeller to the accretion phase.}\label{BPo}
\end{center}
\end{figure}
\begin{figure}[htb]
\begin{center}
\includegraphics[height=5.379cm,width=8.cm]{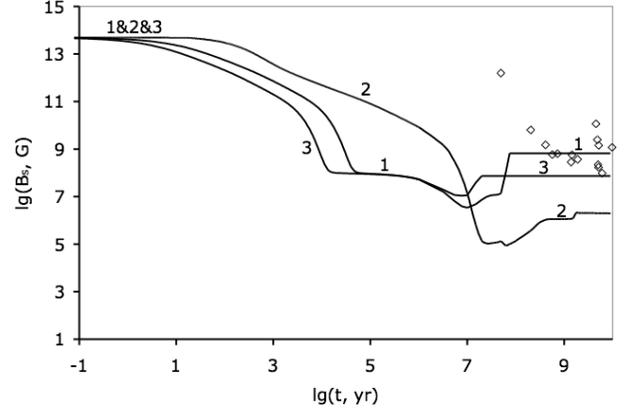} 
\caption{The evolution of the surface magnetic field for: (1)Q=0.3 (2) Q=0.001 (3)Q=1. The initial surface magnetic filed is $B_{si}=5\times10^{13}$ G and the initial core magnetic field is $B_{ci}=10^{11}$ G.
Parameters are the same as in Fig 1.}\label{BQ}
\end{center}
\end{figure}
\begin{figure}[htb]
\begin{center}
\includegraphics[height=5.404cm,width=8cm]{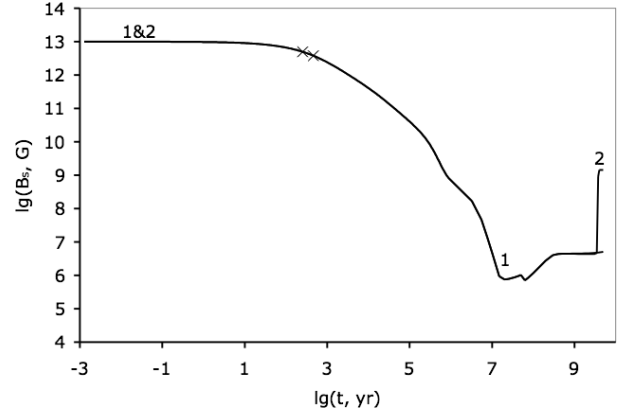} 
\caption{The evolution of the surface magnetic field for a binary neutron star. (1) The orbital evolution is taken into account. (2) The orbital evolution is neglected.  Initially the neutron star is in ejector phase. The first cross corresponds to the point of the transition to the propeller phase and the second cross corresponds to the transition to the accretion phase. Presence of the orbital evolution does not effect the transition points much.
Parameters are: $\dot{M}_2=10^{-14}$ M$_\odot/$yr, $B_{si}=10^{13}$ G, $B_{ci}=10^{11}$ G, $P_{oi}=4$ days, and the rest of the
parameters are the same as in Fig 1.}\label{Ba}
\end{center}
\end{figure}
\begin{figure}[htb]
\begin{center}
\includegraphics[height=5.329cm,width=8cm]{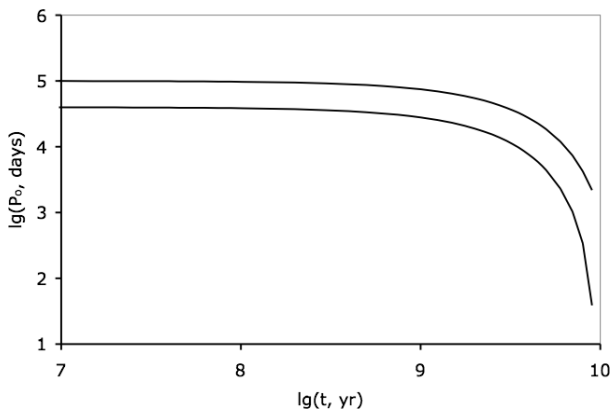} 
\caption{The evolution of the orbital period $P_{o}$ for $P_{oi}=4$ days and $P_{oi}=4.6$ days. 
 $B_{si}=5\times10^{13}$ G, $B_{ci}=10^{11}$ G, $\dot{M}_2=10^{-12}$M$_\odot/$yr, and the rest of the
parameters are the same as in Fig 1.}\label{A}
\end{center}
\end{figure}
\begin{figure}[htb]
\begin{center}
\includegraphics[height=5.262cm,width=8cm]{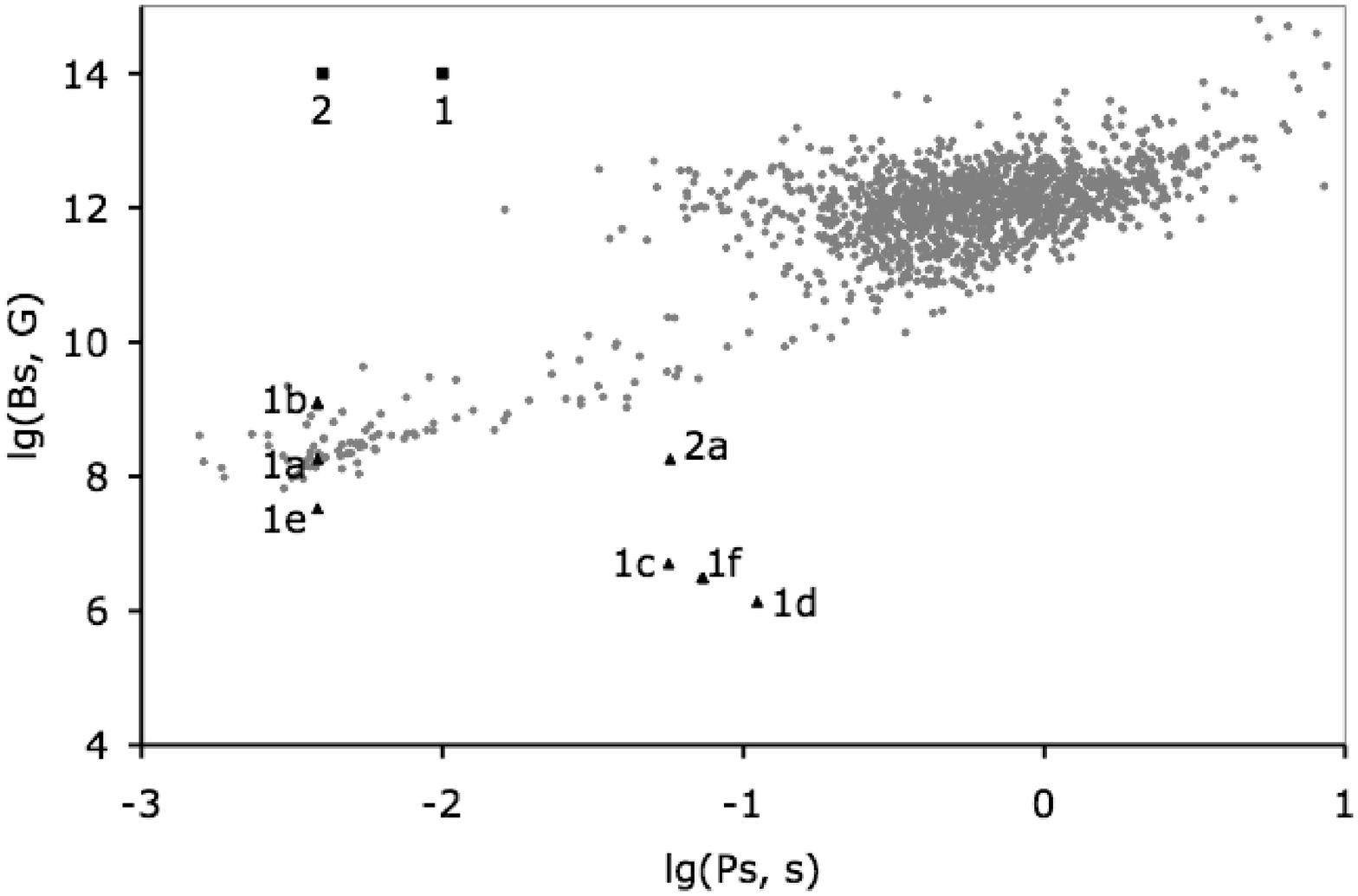} 
\caption{Initial and final state of binary neutron stars in $B-P_s$ diagram for different trial parameters. $5\times 10^{9}$ yrs of the evolution has been considered. Gray circles are observed pulsars. MSPs are located at the lower left region of the diagram (region of the low magnetic field and short spin period). Not all the neutron stars with different parameters end in the MSPs region in $5\times10^9$ yrs. (1) Initial point of the evolution of a young neutron star. $B_{si}=10^{14}$ G, $P_{si}=0.01$ s. (1a) Final state of the evolution of the neutron star (1) for  $\dot{M}_2=10^{-12}$ M$_\odot/$yr, $\zeta=0.1$, $P_{oi}=4$ days. (1b) Final state of the evolution of the neutron star (1) for  $\dot{M}_2=10^{-13}$ M$_\odot/$yr, $\zeta=0.1$, and $P_{oi}=4$ days. (1c) Final state of the evolution of the neutron star (1) for  $\dot{M}_2=10^{-14}$ M$_\odot/$yr, $\zeta=0.1$, and $P_{oi}=4$ days. (1d) Final state of the evolution of the neutron star (1) for  $\dot{M}_2=10^{-15}$ M$_\odot/$ yr, $\zeta=0.1$, and $P_{oi}=4$ days. (1f) Final state of the evolution of the neutron star (1) for  $\dot{M}_2=10^{-12}$ M$_\odot/$yr, $\zeta=0.1$, and $P_{oi}=100$ days. (2)  Initial point of the evolution of a young neutron star. $B_{si}=10^{14}$ G, $P_{si}=0.004$ s. (2a) Final state of the evolution of the neutron star (2) for  $\dot{M}_2=10^{-12}$ M$_\odot/$yr,  $\zeta=0.1$, and $P_{oi}=4$ days. If one considers the evolution of (1) or (2) of a single neutron star the end state is $B_{sf}=10^{9}$ G and $P_{sf}=2.34$ s. Pulsars (2a), (1c), (1f), and (1d) are still in the graveyard at $t=5\times 10^9$ yrs. }\label{BP2}
\end{center}
\end{figure}
We present the results of our numerical calculations for a neutron star of mass $M_n=1.33$M$_\odot$, radius $R_n=7.921$ km, and the radius of the core $R_c=7.43$ km, where we assume that crust extends up to the density $\rho_c=2.4\times 10^{14}$ g cm$^{-3}$. We use the BPS  equation of state \cite{Baym} in the crust of the neutron star. One need to test the results for various equations of states and compare them to the population of the MSPs.
We use the cooling data from Neutron Star Theory Group at UNAM \cite{Dany}. We do not take into account the dependence of the temperature into the densities inside the neutron star. For the further improvement of the model this effect has to be taken into account. Data of the observed pulsars is taken from \cite{AF} (refer to Table III). We assume a very  small value for the moment of the inertia of neutron star $I=10^{40}$ g cm$^{-2}$. Such of the moment of inertia is possible if the center of the neutron star is very dense and compact \cite{Lat} (we have used BPS equation of state only as an approximation in the crust). This small value of the moment of inertia causes the fast rate of change of the spin period of the neutron stars with low mass main sequence companions in the propeller and accretion phases. We show that this scenario is in favor of the observations. Similar values of the moment of inertia have been assumed by other authors \cite{Ur}. However, the main features of this model are independent of the value of the moment of inertia.
Numerical calculations were implemented in $C{++}$ code. We consider some reasonable values for the initial magnetic field of the neutron star, $B_{si}$, intitial rotational period $P_{si}$, the initial orbital period of the binary $P_{oi}$, speed of the stellar wind $V_{w}$, rate of mass loss from the low main sequence companion $\dot M_2$, impurity parameter $Q$, and the efficiency factor $\zeta$ (see TABLE \ref{tab:table 2}). 
We assume that $\dot{M}_2$, $\zeta$, $V_w$ (up to the Roche Lobe phase), and $Q$ are constant parameters of the evolution. 
 For simplicity, we assume that an accretion rate $\dot{M}_2=10^{-14}-10^{-10} $M$_{\odot}$ yr$^{-1}$ heats up the neutron star up to $T_{ac}=3\times 10^6$ K, and an accretion rate $\dot{M}_2=10^{-10}-10^{-9}$ M$_{\odot}$yr$^{-1}$ or higher heats up the neutron star up to $T_{ac}=3\times 10^{8}$ K. 
However, the neutron star might enter the accretion phase at a relatively young age, when its surface temperature is higher than $T_{ac}$. In this case, for simplicity, we let the neutron star cool down freely to the accretion temperature $T_{ac}$, and then keep the star at this temperature as long as the accretion phase continues. 

We considered the evolution of a binary neutron star. Our aim was to find a more realistic model explaining the formation of MSPs, taking into account the effects of the core magnetic field and orbital evolution. The influx of the magnetic field is a unique signature of a binary neutron star with a superconductive core.
In this model the magnetic field and the spin evolution 
play the most important roles. We were able to  track the evolution of a neutron star in the $B_s-P_s$ diagram. The spin, magnetic field, and the orbital evolution are coupled together. 
We consider only the low mass main sequence companions. The evolution has been considered for a period of $5\times 10^{9}$ yrs. The evolution has been stopped before the Roche-Lobe overflow.

 Fig \ref{Bt1} and \ref{Bc} compare the evolution of a binary and a single neutron star, and the effect of the superconductive core. For the single neutron star the magnetic field always decreases. For the binary neutron star if the effect of the core magnetic field is not included the surface magnetic field always decreases. For the binary neutron star with the superconductive core the magnetic field starts to increase at some point during the accretion phase. During the accretion phase the neutron vortices move inward to the core (Fig \ref{V1}). Inward motion of the neutron vortices does not imply the inward motion of the proton fluxoids. To derive the direction of the motion of the fluxoids we have considered various forces which act on the proton fluxoids at the core-crust boundary (Fig \ref{F2} and Fig \ref{F1}). Finally at some point of the evolution fluxoids start to move inward into the core (Fig \ref{V1}). When the fluxoids start moving inward into the core, the core magnetic field increases. $B_c$ gives us the inner boundary condition for equation (\ref{eq1-2}). The increase in the strength of the core magnetic field causes the enhancement of the surface magnetic field of the neutron star. One might wonder if the magnetic field can increase to such a large values that would destroy the standard scenario of the formation of the MSPs. Fortunately, this is not the case. A neutron star does not spin up to arbitrary short spin periods. There is a minimum limit on the spin period of a neutron star \ref{Peq}. 
When this limit reaches the neutron star spin-up phase stops ($\dot{P}_s=0$) temporary as long as $P_s=P_{eq}$ (as long as the neutron star is on the spin-up line). Consequently $v_{n}=0$ as long as $P_s=P_{eq}$. We assume that $v_p=0$ when $v_n=0$. Therefore, the increase of the core magnetic field stops at this point, at least temporarily. 
 If the core magnetic field is kept constant for some period of the time, the surface magnetic field starts decreasing again (refer to the short tail of the Fig \ref{Bs1}). To have another period of the spin-up, the surface magnetic field of the neutron star should decrease. This will cause the decrease of $P_{eq}$. Therefore, the neutron star may enter another short period of the spin up and increase of the magnetic field. In this case the neutron star might spend a very large amount of time without moving much in the $B_s-P_s$ diagram.  
  \begin{figure}[htb]
\begin{center}
\includegraphics[height=5.2544cm,width=8cm]{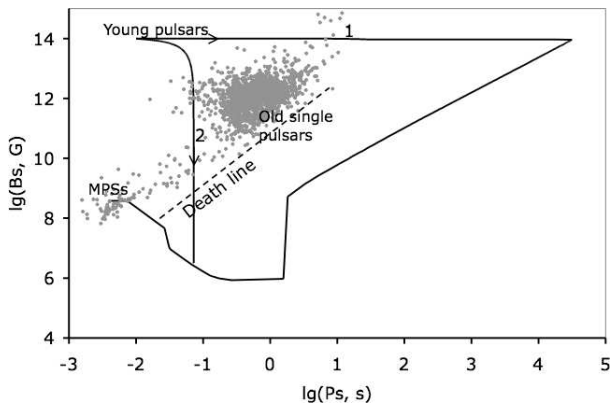} 
\caption{The evolutionary track of the surface magnetic field $B_s$ and spin period $P_s$ for $t=4\times10^9$ yrs. Two different values of the moment of the inertia of the neutron star is chosen. (1) $I=10^{40}$ g cm$^2$ (2) $I=0.3\times10^{45}$ g cm$^2$. The rest of the parameters are the same as Fig 1. }\label{BP5}
\end{center}
\end{figure}
 \begin{table}
\caption{\label{tab:table3}Example of the observed pulsars with low mass companions. Most of these companions are neutron stars (NSs) or white dwarfs (WDs). Unfortunately, there are not many observed pulsars with low mass main sequence (MS) companions. This is not the whole set of observation data we have used. This data for this table is taken from \cite{AF}. We have rounded the numbers here (in figures more accurate numbers are used). }
\begin{ruledtabular}
\begin{tabular}{ccccccc}
Name& $CS~~P_s$&  $P_o$ &  $M_2$& Age& $B_s$\\
$ $ & $ $ $~~~~\mathrm{(s)}$ & $\mathrm{(days)}$ & $$ & $\mathrm{(yrs)}$ & $\mathrm{(G)}$\\ \\
 \hline\\
 $0751+1807$ & $\mathrm{MS}$  $0.003$    &       $0.26$     &  $0.1$  & $7.0\times10^{9}$  &$1.67\times10^{8}$\\
 $J1903+0327$ & $\mathrm{MS}$      $0.002$    &    $95.17$         &      $0.89$ &  $1.8\times10^{9}$ & $2.0\times10^{8}$\\
  $J1913+16$ & $\mathrm{MS}$  $0.05$    &       $0.32$     &  $0.8$  & $1.0\times10^{8}$  &$2.28\times10^{10}$\\
 $B1718-19$ & $\mathrm{MS}$  $1.0~~$    &       $0.2$     &  $0.1$  & $9.8\times10^{6}$  &$1.29\times10^{12}$\\
$J0621+1002$  & $\mathrm{WD}$  $0.03$ & $8.3$ & $0.4$ & $9.6\times10^{9}$ & $1.1\times10^{9}$\\
$J2145+0750$  & $\mathrm{WD}$  $0.01$ & $6.8$ & $0.4$ & $8.5\times10^{9}$ & $7.0\times10^{8}$\\
 $J1435-6100$  & $\mathrm{WD}$  $0.01$ & $1.3$ & $0.9$ & $6.0\times10^{9}$ & $4.8\times10^{8}$\\
$J1157-5112$  & $\mathrm{WD}$     $0.04$ &  $3.5$   &        $1.18$  &$4.8\times10^{9}$ & $2.5\times10^{9}$\\
$J1528-3146$ & $\mathrm{WD}$      $0.06$ &   $ 3.2$      &       $0.9$  & $3.9\times 10^{9}$ & $3.9\times10^{9}$\\
$J1802-2124$ & $\mathrm{WD}$    $ 0.01$ &           $0.7$ &      $0.8$ &  $2.8\times10^{9}$ & $9.7\times10^{8}$\\
$J1454-5846$ & $\mathrm{WD}$   $0.05$ &        $12.4$ &          $0.86$ &  $8.8\times10^{8}$ & $6.2\times 10^{9}$\\
$J0737-3039A$ &  $\mathrm{WD}$  $ 0.02$ &        $ 0.10$   &           $1.3$ & $2.0\times10^{8}$ & $6.4\times10^{9}$\\
$B2303+46$ &  $\mathrm{WD}$ $1.06$     &        $12.3$    &       $1.1$ &  $2.9\times 10^{7}$ & $7.8\times 10^{11}$\\
$J1141-6545$   & $\mathrm{WD}$ $0.39$ &        $0.2$       &   $0.9$&  $1.5\times10^{6}$ & $1.3\times10^{12}$\\
$B1820-11$   & $\mathrm{WD}$ $0.39$ &        $0.2$       &   $1.0$&  $1.4\times10^{6}$ & $1.3\times10^{12}$\\
$J1022+1001$   & $\mathrm{~~~~}$ $0.01$   & $7.80$      &     $0.7$& $6.0\times10^{9}$ &  $8.5\times10^{8}$\\
$B0655+64$   & $\mathrm{~~~~}$ $0.19$   & $1.02$      &     $0.6$& $4.5\times10^{9}$ &  $1.1\times10^{10}$\\
$J1439-5501$   & $\mathrm{~~~~}$ $0.03$   & $2.12$      &     $1.11$& $3.2\times10^{9}$ &  $2.0\times10^{9}$\\
$B1820-11$   & $\mathrm{~~~~}$ $0.09$   & $13.6$      &     $0.4$& $1.5\times10^{9}$ &  $9.7\times10^{9}$\\
$J1753-2240$   & $\mathrm{~~~~}$ $0.27$   & $357$      &     $0.6$& $3.2\times10^{6}$ &  $6.2\times10^{11}$\\
$J1748-2446$   & $\mathrm{~~~~}$ $0.008$   & $0.38$      &     $0.5$& $---$ &  $---$\\
$J1811-1736$ & $\mathrm{NS}$  $0.10$    &       $18.78$     &  $0.9$  & $1.8\times10^{10}$  &$3.1\times10^{9}$\\
$J1829+2456$  & $\mathrm{NS}$   $0.04$    &  $1.18$ &       $1.26$ &  $1.2\times10^{10}$ & $1.5\times10^{9}$\\
$J1756-2251$ & $\mathrm{NS}$     $ 0.03$         &   $0.32$    &     $1.0$ &  $4.4\times 10^{8}$ &  $5.4\times 10^{9}$\\
$B1534+12$ & $\mathrm{NS}$ $0.04$ &    $ 0.42$ &            $1.3$ &$ 2.5\times10^{8}$ & $ 9.7\times10^{9}$\\
$B1913+16 $ & $\mathrm{NS}$   $ 0.06$  &   $0.32$ &          $0.8$ &  $1.1\times10^{8}$ & $2.3\times 10^{10}$\\
$J1906+0746$&  $\mathrm{NS}$   $ 0.14$ &     $0.16$ &       $0.8$&  $1.1\times10^5$ &  $1.7\times 10^{12}$\\
\end{tabular}
\end{ruledtabular}
\end{table}
Pulsars with a weak binary, low rate of mass loss from the companion $\dot{M}_2$, need much more time to enter to the region of the MSPs. Fig \ref{BP} presents the evolution of a neutron star in the $B_s-P_s$ diagram. Pulsar (1) in this figure with a companion that is losing mass at rate $M_2=10^{-14} $M$_\odot $yr$^{-1}$ is very far from the region of the millisecond pulsars after $t=4\times 10^{9}$yrs of the evolution. After $7\times10^{9}$yrs of the evolution this pulsar joins the region of the millisecond pulsars.
 For a binary neutron star the final value of the surface magnetic field and the position of the neutron star in the $B_s-P_s$ diagram is not very sensitive to the initial strength of the core magnetic field or the initial strength of the surface magnetic field (Fig \ref{Bs1}-\ref{BP3}). This is due to the existence of the minimum spin period for a neutron star $P_{eq}$. Neutron stars with different initial strength of the core or surface magnetic field reach $P_{eq}$ at almost the same value of the surface magnetic field (the same value of $P_{eq}$), although at different times. Fig \ref{Pt} shows that two pulsars with different initial surface magnetic fields reach the same value of the spin period. After this point the surface magnetic field remains almost a constant for $10^1-10^{2}$ yrs before it starts decreasing again. This feature is unique to binary neutron stars. For example, the final stage of single pulsars is very sensitive to the initial strength of the core magnetic field (Fig \ref{Bs2}). On the contrary, the value of the final surface magnetic field is very sensitive to the parameters of the binary, such as the rate of the stellar mass loss from the companion $\dot{M}_2$,  the initial orbital period of the binary $P_{oi}$, the speed of the stellar wind $V_w$, and the coefficient factor $\zeta$. Neutron stars with different values of $\dot{M}_2$ reach the minimal spin period $P_{eq}$ at different values of the surface magnetic field (Fig \ref{Bt5}). Therefore, they end up at different values of $B_{sf}$. Fig \ref{BP} shows that also the path of a neutron star in $B-P_s$ diagram strongly depends on $\dot{M}_2$. Fig \ref{BM2} illustrates the dependence of the finial strength of the surface magnetic field to the rate of the mass loss from the companion $\dot{M}_2$ for two different values of the initial orbital period. Note that not all these pulsars have reached their $P_{eq}$ in the period of the evolution we have considered, $t=5\times10^9$yrs. The dashed line marks the most reasonable range of the surface magnetic field of the observed  MS pulsars $B_s=10^7-10^{10}$G. Some of the pulsars which are below this line will gain a stronger surface magnetic field at later stage of their evolution. This feature has been illustrated in Fig \ref{BP3}. This property of these pulsars is due to the low value of $\dot{M}_2$. Namely, these pulsars need more time to reach their minimal spin period $P_{eq}$.  Fig \ref{BPo} illustrates the dependence of the final strength of the surface magnetic field of a neutron star on the initial orbital period of the binary for two different values of the speed of the stellar wind $V_w$. Therefore, this figure also shows that  the evolution depends on $V_w$. Pulsars in binaries with long orbital period need to spend more time in the accretion phase to end up in the region of the millisecond pulsars. The magnetic field increases during this phase and the spin period decreases. Eventually, the spin period reaches its minimal value at high enough magnetic field. At this stage the pulsar reaches the region of the millisecond pulsars. 

The evolution of the surface magnetic field is very sensitive to the different values of $Q$, due to the dependence of the impurity conductivity on this parameter (Fig \ref{BQ}). We have not considered different equations of state and we have assumed a uniform temperature inside the neutron star. However, the dependence of the surface magnetic field on $Q$ implies a sensitive dependence of the evolution on the conductivity. The conductivity depends on the density at different layers inside the neutron star (equation of state) and on the temperature. Therefore, various equations of the state have to be tested. 
 
Fig \ref{Ba} illustrates that the effect of the orbital evolution in very close binaries can be very important at the very late stages of the evolution. Fig \ref{A} shows the evolution of the orbital period of the binary. 

In this model the magnetic field and spin period of the pulsar change substantially. Therefore, binary pulsars  spend some large period of their life in the graveyard (Fig \ref{BP}, \ref{BP3}, and \ref{BP2}). Pulsars with weaker binaries remain for longer periods in the graveyard. Perhaps this accounts for the very low population of the MSPs in the regions belonging to the normal pulsars. We propose that in other models of the evolution where the effect of the superconductive core is neglected could not be as successful as the current model in describing the absence of the binary pulsars from the region of normal pulsars. In these models (\cite{Ur}, \cite{jaa}) magnetic filed always decreases and final value of the magnetic field should be between $10^7$G and $10^{10}$G. If pulsars in such models get to very low values of the surface magnetic field they can not come back to the region of MSPs. 

In Fig \ref{BP5} we compare the evolution of two pulsars of small and large moments of inertia. Pulsar (2), with large value of the moment of inertia, enters the propeller phase at age $3\times 10^{4}$ yrs, and later enters the accretion phase at age $9\times 10^{5}$ yrs. However, due to the very large value of the moment of inertia its spin does not change much. 
Therefore, this pulsar follows almost vertical paths in the $B_s-P_s$ diagram. However, this pulsar should be visible at the propeller and accretion phases. If this were true, we would observe many binary pulsars at different ages, passing through the region of normal pulsars. Depending on the different values of their spin period and magnetic field, we would find them at different paths in $B_s-P_s$ diagram. Almost all of these pulsars would pass almost vertically through the region of normal pulsars. According to the observations, there are only a handful of binary pulsars in the region of normal pulsars (only $2\%$). Therefore, one needs an evolution that explains the absence of binary pulsars from the region of normal pulsars. This type of the evolution can be provided by low values of the moment of inertia. 
\begin{acknowledgments}
The author wishes to thank Don Page, Sharon Morsink, Andrey Shoom, and Craig Heink for valuable discussions. This research was supported in part by Natural Science and Engineering Research Council of Canada. 
\end{acknowledgments}

\end{document}